%% file: main.tex
\documentclass[letterpaper,twocolumn,10pt]{article}

\newcommand{\system}{Vilamb\xspace}
\newcommand{\sync}{Pangolin\xspace}
\newcommand{\baseline}{No-Redundancy\xspace}

\newcommand{\fio}{fio\xspace}
\newcommand{\redundancy}{system-redundancy\xspace}
\newcommand{\Redundancy}{System-redundancy\xspace}
\newcommand{\paperTitle}{\system: Low Overhead Asynchronous Redundancy 
	for Direct Access NVM}

\usepackage{usenix2019_v3}
\usepackage{tikz}
\usepackage{amsmath}
\usepackage{comment}
\usepackage{times}
\usepackage{fullpage}
\usepackage{titlesec}
\usepackage{color}
\usepackage{amsmath}
\usepackage{etoolbox}
\usepackage{graphicx}
\usepackage{subfigmat}
\usepackage{float}
\usepackage{xspace}
\usepackage{makecell}
\usepackage{pifont}
\usepackage{enumitem}
\newcommand{\cmark}{\ding{51}}
\usepackage[linesnumbered,algoruled,boxed,lined]{algorithm2e}
\usepackage[thicklines]{cancel}

\microtypecontext{spacing=nonfrench}

\usepackage{verbatim}

\usepackage[scaled]{inconsolata}

\usepackage[font={small},labelfont=bf]{caption}
\usepackage[capitalize,nameinlink]{cleveref}
\crefname{section}{\S}{\S\S}

\renewcommand{\check}[1]{}
\newif\ifcommenton
\commentontrue
\ifcommenton
\newcommand{\rajat}[1]{{\color{magenta}{\sf [Rajat: #1]}}}
\renewcommand{\check}[1]{{\color{magenta}{\sf [Check: #1]}}}
\newcommand{\andy}[1]{{\color{red}{\sf [Andy: #1]}}}
\newcommand{\greg}[1]{{\color{blue}{\sf [Greg: #1]}}}
\else
\newcommand{\rajat}[1]{}
\newcommand{\andy}[1]{}
\newcommand{\greg}[1]{}
\fi 

\hypersetup{
    colorlinks=false,
    pdfborder={ 0 0 0 }
}

\titlespacing{\section}{0pt}{4pt}{4pt}
\titlespacing{\subsection}{0pt}{4pt}{3pt}
\titlespacing{\subsubsection}{0pt}{0pt}{0pt}
\begin{document}

\date{}

\title{\Large \bf \paperTitle}

\author{Rajat Kateja, Andy Pavlo, Gregory R. Ganger\\rkateja@cmu.edu, pavlo@cs.cmu.edu, ganger@ece.cmu.edu\\Carnegie Mellon University}

\maketitle

\input{abstract}
\input{intro}

\input{background}

\input{design}

\input{eval}

\input{reliability-eval}
\input{conclusion}
\clearpage
\newpage

\bibliographystyle{plain}
\bibliography{references}

\end{document}

%% file: abstract.tex
\begin{abstract}
	\system provides efficient asynchronous \redundancy 
	for direct access (DAX) non-volatile memory (NVM) storage. 
	Production storage deployments often use \redundancy in 
	form of page checksums and cross-page parity.
	State-of-the-art solutions for maintaining \redundancy 
	for DAX NVM either incur a high performance overhead or require 
	specialized hardware. 
	The \system user-space library maintains \redundancy 
	with low overhead
	by delaying and amortizing the \redundancy updates
	over multiple data writes. 
	As a result, \system provides 3--5$\times$ the throughput of
        the state-of-the-art software solution at high operation rates.
	For applications that need \redundancy with high performance, 
	and can tolerate some delaying of data redundancy, 
	\system provides a tunable knob between performance and quicker 
	redundancy. Even with the delayed coverage, 
	\system increases the mean time to 
	data loss due to firmware-induced corruptions by up to two orders of magnitude in comparison to 
	maintaining no \redundancy. 
\end{abstract}

%% file: intro.tex
\section{Introduction}
\label{sec:intro}
Non-volatile memory (NVM)
storage combines DRAM-like access 
latencies and granularities with disk-like 
durability~\cite{pcm-isca, pcm-wear-leveling-isca, 3dxp, memristor, rio-asplos}.
Direct access (DAX) 
to NVM data exposes raw NVM performance to 
applications. 
Applications using DAX map NVM files
into their address spaces and access data with load and store instructions, 
eliminating system software overheads associated with conventional 
storage interfaces.

\nocite{bpfs-sosp, pmfs-eurosys, nova-fast, soupfs-atc, nova-fortis-sosp}
\nocite{peloton, nstore-sigmod, foedus-sigmod, wbl-vldb} 
\nocite{pmemcached-hotstorage, redis-pmem}

Production storage demands fault tolerance in addition to 
non-volatility and performance. 
Whereas some fault tolerance mechanisms extend to DAX NVM storage 
trivially (e.g., background scrubbing), 
others
do not. 
In particular, mechanisms for resilience against 
device-firmware-bug-induced data corruption 
fit poorly. 
FS-level page checksums enable detection 
of firmware-bug-induced data corruption, and cross-page 
redundancy enables recovery from such 
corruptions~\cite{ironfs-sosp, data-corruption-tos, 
storage-subsystem-failures-tos, 
long-term-digital-storage-eurosys, 
data-integrity-techniques-applications-storagess}.
We use \emph{\redundancy} to 
refer to FS level checksums and cross-page redundancy. 

Maintaining \redundancy for DAX NVM storage,
without forfeiting its performance benefits, is challenging for two reasons.
First, accesses via load and store instructions bypass system
software, 
removing the straightforward ability to detect and act on data changes
(e.g., to update \redundancy).
Second, NVM's cache-line granular writes increase the overhead
of updating \redundancy (e.g., checksums) that 
is usually computed over sizeable data regions (e.g., pages)
for effectiveness and space efficiency.

\nocite{ironfs-sosp, zfs-end-to-end-integrity-fast, gfs-sosp}

The state-of-the-art solution for DAX NVM \redundancy 
is the Pangolin library~\cite{pangolin-atc}. 
Pangolin addresses the challenge of  
system software bypass by requiring applications to use its 
transactional API. This enables Pangolin to mediate 
and act on data accesses 
To address the incongruence in DAX write and \redundancy
granularities, Pangolin introduces micro-buffering and 
per-object checksums. Pangolin buffers application 
writes in DRAM and updates the NVM only on transaction 
commits. This buffering also enables Pangolin to use data diffs
to make \redundancy updates more efficient.

\begin{figure}[t]
	\center{\includegraphics[width=.7\columnwidth, keepaspectratio]{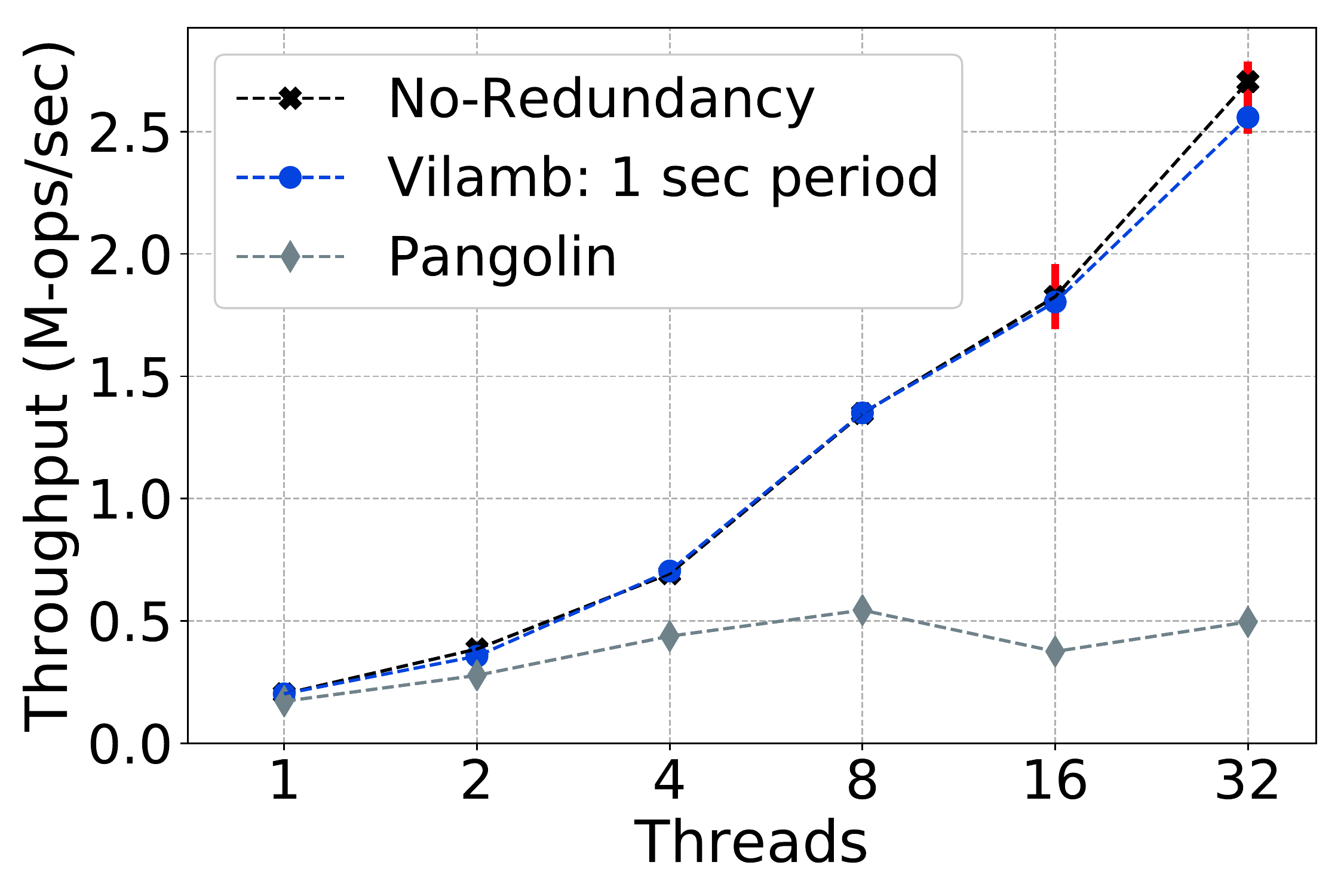}}
\caption{Throughput for a PMDK key-value store
when using three \redundancy options,
as a function of the number of threads
performing PMDK's insert-only benchmark workload.
(Details in \cref{sec:kvs}; RBtree results shown here.)
}
\label{fig:pangolin-overhead}
\end{figure}

Even with Pangolin's well-optimized design,
synchronous \redundancy updates incur significant overhead.
For example, \cref{fig:pangolin-overhead}
shows that Pangolin reduces key-value insert throughput by 10--20\%
at low insert rates, compared to a \baseline baseline,
and by up to 80\% at high rates.
Fundamentally, any software-based 
synchronous 
approach will struggle
with high throughput updates because it must update \redundancy 
on every operation.
A recently proposed specialized hardware controller offers 
low-overhead synchronous DAX NVM \redundancy~\cite{tvarak-pdltr},
but it is unlikely to be available in systems soon. 

This paper describes \textbf{\system}, a user-space 
library for efficient asynchronous DAX NVM \redundancy.
\system moves \redundancy updates out of the critical path 
and delays them to 
amortize the overhead over multiple data updates. 
Delaying the \redundancy updates creates a configurable 
trade-off between the delay before 
updated data is covered and performance.
\cref{fig:pangolin-overhead} shows that updating 
\redundancy every second with \system 
reduces the \baseline throughput by only 6\%, even at the highest
throughput level; this corresponds to 5$\times$ higher 
throughput than Pangolin. Although \system leaves a fraction of data 
briefly uncovered, it increases the mean time 
to data loss (MTTDL) due to firmware-induced corruptions 
by 112$\times$ over \baseline for this benchmark.


Unlike \sync, \system does not require 
applications to adopt a particular access interface  
to identify data updates. 
Instead, \system repurposes page table dirty bits to
efficiently identify of data updates. \system 
marks pages with updated \redundancy as clean 
and identifies pages with outdated \redundancy  
by checking their dirty bit.
We implement a kernel module that \system uses for 
batched fetching and clearing for dirty bits.
\system ensures atomic and consistent \redundancy 
updates for all dirty pages by 
using shadow copies of 
dirty bits and 
leveraging batteries that are 
common in 
production environments~\cite{no-compromises-sosp, 
fasst-osdi, viyojit-isca, cdn-batteries-socc, 
ups-isca, underprovision-ups-asplos, 
distributed-ups-isca,datacenter-energy-storage-sigmetrics}. 



Extensive evaluation with eight macro- and micro-benchmarks 
demonstrate \system's efficacy.
\system with a 1~sec delay between 
\redundancy updates reduces single-threaded Redis' YCSB throughput 
by only 1.6--17\%, compared to 13--18\% for Pangolin.
Increasing the delay to 10 seconds further
reduces \system's overhead to 0.1--6\%. 
Similar to \cref{fig:pangolin-overhead}, 
\system offers 3--5$\times$ higher throughput
than \sync at high insert rates for 
all five of Intel's PMDK key-value stores.
By protecting the clean pages
from firmware-bug-induced corruption, 
\system increases the MTTDL 
over \baseline. For example, 
\system with a 1~sec \redundancy update period 
increases Redis' MTTDL 
by 15$\times$ and 74$\times$ over \baseline for a 
write-heavy and ready-heavy YCSB workload, respectively.
Detailed timing breakdowns with \fio microbenchmarks 
and battery cost analysis confirm \system{}'s design decisions. 


\nocite{redis-pmem, pmemcached-hotstorage, whisper-asplos, nvheaps-asplos, 
mnemosyne-asplos, pmemlib, bpfs-sosp, pmfs-eurosys, nova-fast, wbl-vldb}

This paper makes three primary contributions.
First, it identifies asynchronous \redundancy as
an important addition to the toolbox of DAX NVM \redundancy 
solutions.
Second, it describes \system's efficient delayed 
\redundancy design that improves performance for 
applications that can tolerate delayed coverage.
Third, it quantifies \system{}'s efficacy, cost, and 
reliability via extensive evaluation with
eight macro- and micro-benchmarks. 

%% file: background.tex
\section{Background and Related Work}
\label{sec:background}

This section provides background 
on direct-access (DAX) NVM and \redundancy,
and the challenges that DAX poses for maintaining \redundancy.
It then describes the solution space and how \system and related
work fit into it.

\subsection{Direct-Access (DAX) NVM}
\label{sec:nvm-background}
NVM refers to a class of memory 
technologies that have access latencies comparable 
to DRAM
and that retain their contents across power outages like disks.
Various NVM technologies, 
such as 3D-XPoint~\cite{3dxp, nvm-perf-arxiv}, Memristors~\cite{memristor}, 
PCM~\cite{pcm-isca, pcm-wear-leveling-isca}, and
battery-backed DRAM~\cite{rio-asplos, saferam-vldb}, are either 
already in-use or expected to be available soon. 
In this paper, we focus on NVM that is accessible like DRAM DIMMs
rather than like a disk~\cite{optane-ssd}. That is, 
NVM that resides on the memory bus,
with load/store accessible data that 
moves between CPU caches 
and NVM at a cache-line granularity. 
Although applications can continue to access 
NVM via conventional FS 
interface, 
doing so incurs the overhead of system calls, 
and (potentially) data copying and 
inefficient general-purpose file system 
code~\cite{bpfs-sosp, nova-fast,
pmfs-eurosys, scmfs-sc, aerie-eurosys, splitfs-sosp}. 

The DAX interface to NVM eliminates system software 
overheads, enabling applications to leverage raw NVM 
performance. 
With DAX, applications map NVM pages into their address
spaces and access persistent data via load and store instructions.
File systems that map a NVM file into the application 
address space (bypassing the page cache) on a \texttt{mmap} 
system call are referred to as DAX file systems 
and said to support DAX-mmap~\cite{linux-dax, pmfs-eurosys, 
nova-fortis-sosp}. 
DAX is widely used for adding persistence to conventionally 
volatile in-memory DBMSs~\cite{pmemcached-hotstorage, 
redis-pmem,mojim-asplos,pmse} and is poised as the 
``killer use-case'' for NVM.

DAX-mmap helps applications realize NVM performance benefits, 
but requires careful reasoning to ensure data consistency. 
Volatile processor caches can write-back data in arbitrary order, 
forcing applications to use 
cache-line flushes  
and memory fences for durability and ordering.
Transactional NVM access libraries ease this burden by 
exposing simple transactional APIs to applications and 
ensuring consistency on their 
behalf~\cite{pmemlib, nvheaps-asplos, mnemosyne-asplos, 
lsnvm-atc, oracle-nvm-direct}.
Alternatively, the system can be equipped with enough battery to allow
flushing of cached writes to NVM before a power failure~\cite{wsp-asplos, kiln-micro,
pcommit-deprecated}; our work assumes this option. 

\begin{table*}[t]
    \centering
    \input{tables/solutions.tex}

    \caption{Solutions for DAX NVM \redundancy and their trade-offs.}
    \label{table:design-space}
\end{table*}

\subsection{System-Redundancy}
\label{sec:system-redundancy}
Many production storage systems implement \redundancy, 
in the form of FS level page checksums and cross-page redundancy, 
to protect against firmware-bug-induced data 
corruption~\cite{latent-sector-errors-tos, 
ironfs-sosp, zfs-end-to-end-integrity-fast, gfs-sosp}.
Device firmwares are susceptible to bugs, like any software, because of 
their complex functionalities, such as address translation
and wear leveling. 
A class of these bugs, namely lost write bugs and 
misdirected read or write bugs, can cause data 
corruption~\cite{ironfs-sosp, data-corruption-tos, 
storage-subsystem-failures-tos, 
long-term-digital-storage-eurosys, 
data-integrity-techniques-applications-storagess}. 
Lost write bugs cause the firmware 
to incorrectly consider a write as completed 
without actually writing the data on to the device media. 
Misdirected read or write bugs cause the firmware to 
access (read or write) data at a wrong location on 
the device media. 

Firmware bugs can corrupt data that an application 
is actively accessing 
as well as data at rest. 
An example of a firmware bug affecting actively accessed 
data would be a misdirected read bug that causes 
the firmware to return incorrect data for an 
application read. 
On the other hand, lost write or address mapping bugs that are
triggered when the firmware 
is performing wear-leveling could corrupt 
data at rest.

Storage systems can detect and recover from 
firmware-bug-induced corruption using 
\redundancy~\cite{ironfs-sosp, zfs-end-to-end-integrity-fast, 
pvfs-checksums-pdsw}. 
For example, a FS can store and access page checksums separately 
from the data, making it unlikely for a firmware 
bug to affect both the data and its FS-level checksum in the same manner. 
An FS-level  checksum mismatch can then flag firmware-bug-induced 
corruption, which the FS can recover from by 
using cross-page parity. 

Many storage systems implement \redundancy in addition 
to a variety of other fault-tolerance 
mechanisms~\cite{designing-for-disasters-fast, 
seneca-atc, snapmirror-fast, gfs-sosp, nova-fortis-sosp, 
zfs-end-to-end-integrity-fast, btrfs-tos, wafl-usenix, 
wafl-integrity-fast}. 
In particular, storage systems implement \redundancy 
even in the presence of device-level error 
correcting codes 
(ECCs)~\cite{nvm-ecc-micro, ssd-error-characterization-ieee, 
nand-ecc-isca}.
ECCs are designed for, 
and effective against, random bit flip induced corruption.
However, they are ineffective 
against most firmware-bug-induced corruption, because 
they are computed, stored, and accessed as a single unit
with the data at a very low level of the device's firmware or hardware.

\subsection{System-Redundancy for DAX NVM}
\label{sec:challenges}
Production NVM storage deployments
will require similar levels of fault-tolerance 
as conventional storage 
deployments, including \redundancy. 
Unsurprisingly, recently proposed NVM storage system designs include 
\redundancy~\cite{nova-fortis-sosp, plexistore-nvmw, 
pangolin-atc, tvarak-pdltr}. Among these proposals, 
file systems like Nova-Fortis~\cite{nova-fortis-sosp}
and Plexistore~\cite{plexistore-nvmw} implement 
\redundancy only for data that is accessed via 
the FS interface. 

Maintaining \redundancy for DAX NVM
is challenging for two reasons:
(i)~hardware controlled data movement,
and (ii)~cache-line granular writes.
\\
\textbf{Hardware Controlled Data Movement}:
Applications' data writes to DAX NVM bypass 
system software. This lack of software control 
makes it challenging for the storage software
to identify updated NVM pages for which 
it needs to update \redundancy.
\\
\textbf{Cache-line Granular Writes}: 
Incongruence in the size of 
DAX writes and the size of pages over which 
\redundancy is usually maintained 
increases the overhead of
maintaining \redundancy. 
Most storage systems maintain
\redundancy over sizeable blocks (e.g., 4K page checksums)
for space efficiency. 
Cache-line granular writes require reading (at least) an 
entire page to update the \redundancy. 
Whereas RAID systems solve a similar ``small write'' problem 
by reading the data before updating 
it~\cite{raid-sigmod}, a DAX NVM storage system software cannot 
use this 
solution. As discussed above, direct access to NVM bypasses
system software, prohibiting the use of pre-write values
for incremental \redundancy updates.

\subsection{Related Work: Solution Design Space}
\label{sec:solutions}
\cref{table:design-space} summarizes the
design space of DAX NVM \redundancy solutions and 
the tradeoffs among the three 
options (including \system) in
the toolbox.

\textbf{Pangolin}~\cite{pangolin-atc} is a user-space library 
that maintains DAX NVM \redundancy synchronously 
by requiring applications to explicitly inform it about 
their data updates;
applications piggyback these notifications 
on Pangolin's transactional interface. 
Pangolin offers strong coverage 
(immediate \redundancy updates and verification) and 
does not require any specialized hardware resources 
(because it is a software-based solution). 
Pangolin addresses the mismatch of fine-grained 
DAX updates with large checksum ranges by requiring explicit object
definitions and maintaining per-object checksums 
instead of per-page checksums.

Pangolin is well-tuned, including several overhead-reducing mechanisms, 
making it the state-of-the-art for an in-line software-only solution.
Yet, Pangolin still incurs significant performance overhead 
(up to 80\%) in many cases.
Fundamentally, Pangolin's synchronous \redundancy update 
design requires updating \redundancy at the same rate 
at which an object is being modified; this becomes 
costly for the high update rates enabled by NVM. 
Pangolin's per-object 
checksums also incur higher space overhead 
for small data objects.
Also, importantly, 
Pangolin only works for applications that can be and are modified
to use its object-based transactional interface. 
Applications that manage NVM data themselves using other data models, 
such as NVM-optimized databases~\cite{nstore-sigmod}, 
may not be easily fit to Pangolin's interface.

\textbf{Tvarak}~\cite{tvarak-pdltr} is a hardware controller co-located with the 
last level cache (LLC) that the FS can offload \redundancy 
maintenance work onto. Tvarak is able to 
identify data updates by the virtue of being interposed 
in the data path. Tvarak offers synchronous \redundancy updates 
and verification, does not restrict applications 
to any specific library/API, and is low-overhead.
However, it requires specialized hardware resources, including 
a controller, on-controller cache, and shared LLC 
partitions. The need for dedicated (and newly proposed) 
hardware resources implies that Tvarak is not available 
for immediate use, and may not be part of commodity servers for many years.
Further, Tvarak introduces cache-line granular checksums for 
DAX-mapped data, increasing the space overhead. 

Prioritizing strong coverage at the expense of 
performance and a restrictive programming model (with Pangolin~\cite{pangolin-atc}), 
or cost and near-term  availability (with Tvarak~\cite{tvarak-pdltr}),
will not be the preferred choice for all applications.
Many applications prioritize performance and use storage systems
wherein some of the fault-tolerance mechanisms (e.g., remote replication
or even persistence) are asynchronous---the fault-tolerance
is still desired, and the more coverage the better, but not at a high
performance cost~\cite{dynamodb-sosp, 
snapmirror-fast, designing-for-disasters-fast, 
seneca-atc}. 


\textbf{\system} is a software library that
embraces an asynchronous approach to updating \redundancy 
for updated data. 
Like other asynchronous redundancy-update approaches, it identifies
and completes required \redundancy updates in the background.
Indeed, it does both aspects (identifying and updating) outside the
critical path of application accesses.
As such, \system can provide low-overhead DAX NVM \redundancy.
Also, \system does not impose any programming 
model restrictions and does not require any specialized
hardware resources.
But, \system reduces the data coverage guarantees by delaying 
\redundancy updates.
Specifically, recently modified pages may not be covered when a
firmware bug affects them.
So, \system can be a good option when applications 
desire high performance and/or are not a good fit 
for Pangolin-like API. 
and view partial \redundancy
coverage is as better than none.

%% file: tables/solutions.tex
%
%
{\small
\setlength{\tabcolsep}{3pt}

\definecolor{mygreen}{rgb}{0.0, 0.7, 0.0}

\begin{tabular}{r||c|c|c|c}
	\makecell[r]{\textbf{Solution}} & 
	\makecell[c]{\textbf{Coverage}\\\textbf{Guarantees}} & 
	\makecell[c]{\textbf{Performance}\\\textbf{Overhead}} & 
	\makecell[c]{\textbf{Programming}\\\textbf{Model}}  &
	\makecell[c]{\textbf{Specialized Hardware}\\\textbf{Requirement}} \\ 
\hline
	\makecell[r]{Pangolin~\cite{pangolin-atc}} & 
	\makecell[c]{\color{mygreen}{Strong}} & 
	\makecell[c]{\color{red}{Medium-to-High}} & 
	\makecell[c]{\color{red}{Restrictive}}  &
	\makecell[c]{\color{mygreen}{None}} \\
 \hline
	\makecell[r]{Tvarak~\cite{tvarak-pdltr}} & 
	\makecell[c]{\color{mygreen}{Strong}} & 
	\makecell[c]{\color{mygreen}{Negligible}} & 
	\makecell[c]{\color{mygreen}{Non-Restrictive}}  &
	\makecell[c]{\color{red}{Yes}} \\
 \hline 
	\makecell[r]{\system} & 
	\makecell[c]{\color{blue}{Configurable}} & 
	\makecell[c]{\color{blue}{Configurable}} & 
	\makecell[c]{\color{mygreen}{Non-Restrictive}}  &
	\makecell[c]{\color{mygreen}{None}} \\
\end{tabular}
}

%% file: design.tex
\section{\system Design and Implementation}
\label{sec:design}
This section begins by describing 
\system's design elements:
delayed \redundancy updates
and repurposing of dirty bits. 
It then describes the effect of \system's 
design on 
resilience against different failures 
and ends with \system's implementation details. 

\subsection{Asynchronous System-Redundancy}
\label{sec:lazy-csum-parity}
\system asynchronously maintains per-page checksums 
and cross-page parity for DAX NVM storage. 
A background thread periodically updates
\redundancy for pages which have been written to 
since \system last updated their \redundancy.
By delaying \redundancy updates, \system amortizes
the overhead over multiple cache-line 
writes to the same DAX NVM page. 

\begin{figure}[t]
	{\includegraphics[width=\columnwidth]{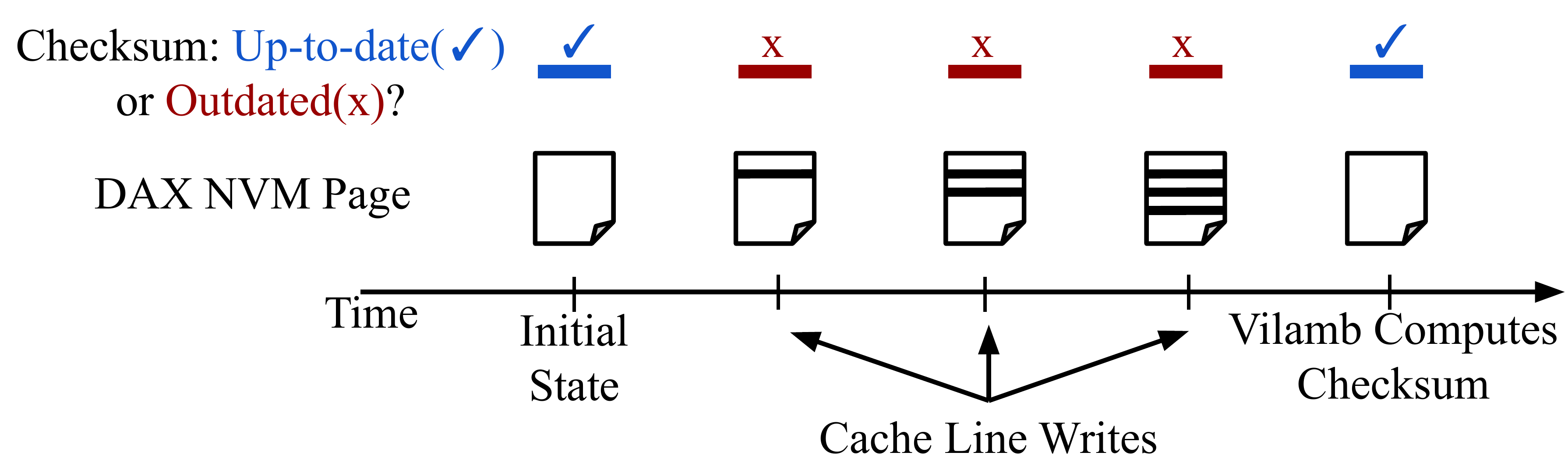}}
\caption{\textbf{Delayed Checksum Computation Example} -- By computing 
per-page checksums asynchronously, \system amortizes the computation overhead 
over multiple cache-line writes to the same NVM page.}
\label{fig:lazy-csum-example}
\end{figure}

\cref{fig:lazy-csum-example} illustrates how \system
reduces work for 
per-page checksums (cross-page parity is not shown in the example, 
but is updated at the same time as the page checksum). 
The figure shows a DAX NVM page and its checksum; 
the checksum can either be up-to-date~(\textcolor{blue}{\cmark}) or
outdated~(\textcolor{red}{x}). In the initial state, the 
checksum is up-to-date with the data. The 
first write to the page makes the checksum stale. 
Instead of updating the checksum immediately, 
\system delays the update until after two more writes. 
By delaying the update \system performs a single 
checksum (and parity, not shown in the figure) computation, instead of three.

\system scrubs the data using a separate background 
thread to detect data corruption. Upon mismatch 
between the page data and checksum for a clean 
page, \system raises an error and halts the program. 
The OS can recover corrupted pages using the 
parity pages, with potential re-mapping 
to different physical pages~\cite{nova-fortis-sosp, pangolin-atc}.

\subsection{Repurposing Dirty Bits}
\label{sec:repurposing-dirty-bits}
The conventional use-case of dirty bits is 
irrelevant for DAX NVM pages, making them 
available for repurposing. 
The dirty bit is conventionally used to identify
updated, or ``dirtied'', in-memory pages that the 
storage system needs to write back 
to persistent storage. 
In case of DAX NVM storage, the file system maps 
NVM-resident files into application address spaces
using the virtual memory system~\cite{pmfs-eurosys, 
linux-dax}. Consequently, even though each mapped 
page has a corresponding dirty bit, 
the conventional semantic of these dirty bits is irrelevant 
because the pages already reside in persistent NVM 
storage.

\system repurposes dirty bits to identify  
pages that have been written to since \system last updated 
their \redundancy. 
When a file is first DAX
mapped, its pages' dirty bits 
are clear and \redundancy  
is up-to-date (potentially updated
during initialization for newly 
created files). 
A page write, which causes its \redundancy to become 
stale, sets the page's dirty bit. In 
each successive invocation, \system's 
background thread updates the \redundancy only for 
pages with their dirty bit set and then clears the
corresponding dirty bits again. 

\textbf{Shadow Dirty Bits}: 
\system carefully orchestrates the non-atomic two-step 
process of updating a page's \redundancy
and clearing its dirty bit;
performing these steps without any safeguard 
is incorrect. 
Clearing the
dirty bit after updating the \redundancy is incorrect
because an interleaved application access can invalidate 
the \redundancy. 
Reversing the order is not safe either.
A checksum verification (e.g., in a scrubbing thread) after 
the dirty bit is cleared, but before the checksum is updated,
would cause a spurious checksum-mismatch. 
\system makes a persistent shadow copy of the dirty bit before clearing it, and 
clears this shadow copy only after completing the redundancy update.
If either of the dirty bit or
its shadow copy is set
for a page, \system knows that the page's redundancy is outdated. 

\subsection{Failure Coverage}
\system's asynchronous approach to \redundancy 
introduces a tunable window of vulnerability. 
Pages that an application writes to remain 
susceptible to corruption until \system 
updates their \redundancy. We describe the 
implication of this window of vulnerability 
for different kinds of failures below.

\textbf{Page Corruption}:
\Redundancy's primary goal is to protect data from 
firmware-bug-induced corruption. 
Additionally, \redundancy also protects from 
random bit flip induced corruptions, though
on-device ECCs are already expected to address those.
\system's delayed checksums would detect corruption to all but 
recently written (dirty) pages. We illustrate  
this 
with an example lost write bug 
triggered in three different scenarios. 

Consider a firmware that uses an on-device write-back 
cache and that suffers from a bug
wherein the firmware 
(infrequently) ``forgets'' to destage some 
data from the cache to the device media.
(1) For the first scenario, consider an application write 
that is evicted 
from the CPU caches to the 
NVM device, is stored in the on-device write-back 
cache, and then lost by the firmware before 
\system updates the corresponding 
page's checksum. 
This would lead to a silent 
corruption
because \system would use the 
incorrect (old) data to compute the checksum. 
(2) For the second scenario, consider that \system updates the 
page's checksum before the firmware bug is triggered 
(i.e., while the data is in the CPU caches or 
in the on-device cache). \system would update the 
checksum correctly in this scenario and 
detect the subsequent corruption because of a data checksum mismatch 
at a later point.
(3) For the third scenario,
imagine the bug affects a clean page while the 
firmware is performing wear leveling. \system would be 
able to detect this data loss in its scrubbing thread.

Among the pages that \system detects as corrupted, 
\system can recover those that belong to stripes with 
all clean pages (and hence, an up-to-date parity). Any  
dirty page in a stripe invalidates the parity. Thus, even 
if the corrupted page is itself clean, \system can recover 
it only if all other pages in its stripe are also clean. 

\textbf{Power Failures}:
\system avoids any inconsistencies between 
data and its \redundancy by ensuring that the \redundancy is made
up-to-date if there is a power failure.
To that end, 
\system leverages battery backups that are common in production 
environments~\cite{no-compromises-sosp, fasst-osdi, viyojit-isca, 
wafl-usenix, ups-isca, underprovision-ups-asplos, distributed-ups-isca}.
Conventional storage systems use batteries to flush 
DRAM to a persistent medium upon a power 
failure~\cite{no-compromises-sosp, fasst-osdi, viyojit-isca, 
wafl-usenix}. 
NVM does not need batteries to make its contents persistent,
because they are already persistent. 
\system instead leverages the battery backup to 
update \redundancy upon a power failure,
ensuring that no pages are left uncovered.
Given that batteries are also used to address other issues, including
brief power losses and spikes~\cite{cdn-batteries-socc}, 
we believe that \system can exploit them for 
updating \redundancy.


\textbf{NVM DIMM Failures or Machine Failures}:
\system's \redundancy is not intended for protection against DIMM 
or machine failures; the storage system can protect against 
these using remote replication~\cite{mojim-asplos, hotpot-socc}.
Being a machine-local fault-tolerance mechanism, \redundancy, 
independent of its implementation, is ineffective against machine 
failures. 
For DIMM failures, \system's asynchronous \redundancy design makes it 
unable to reconstruct the fraction of the pages in the failed DIMM that 
belonged to a stripe with outdated \redundancy. 
Although the storage system could still recover a large fraction 
of the data (\cref{sec:reliability-analysis}), it would need other redundancy 
to recover the remaining data. 

\begin{figure}[t]
	\centering
  \includegraphics[width=0.8\columnwidth, keepaspectratio]{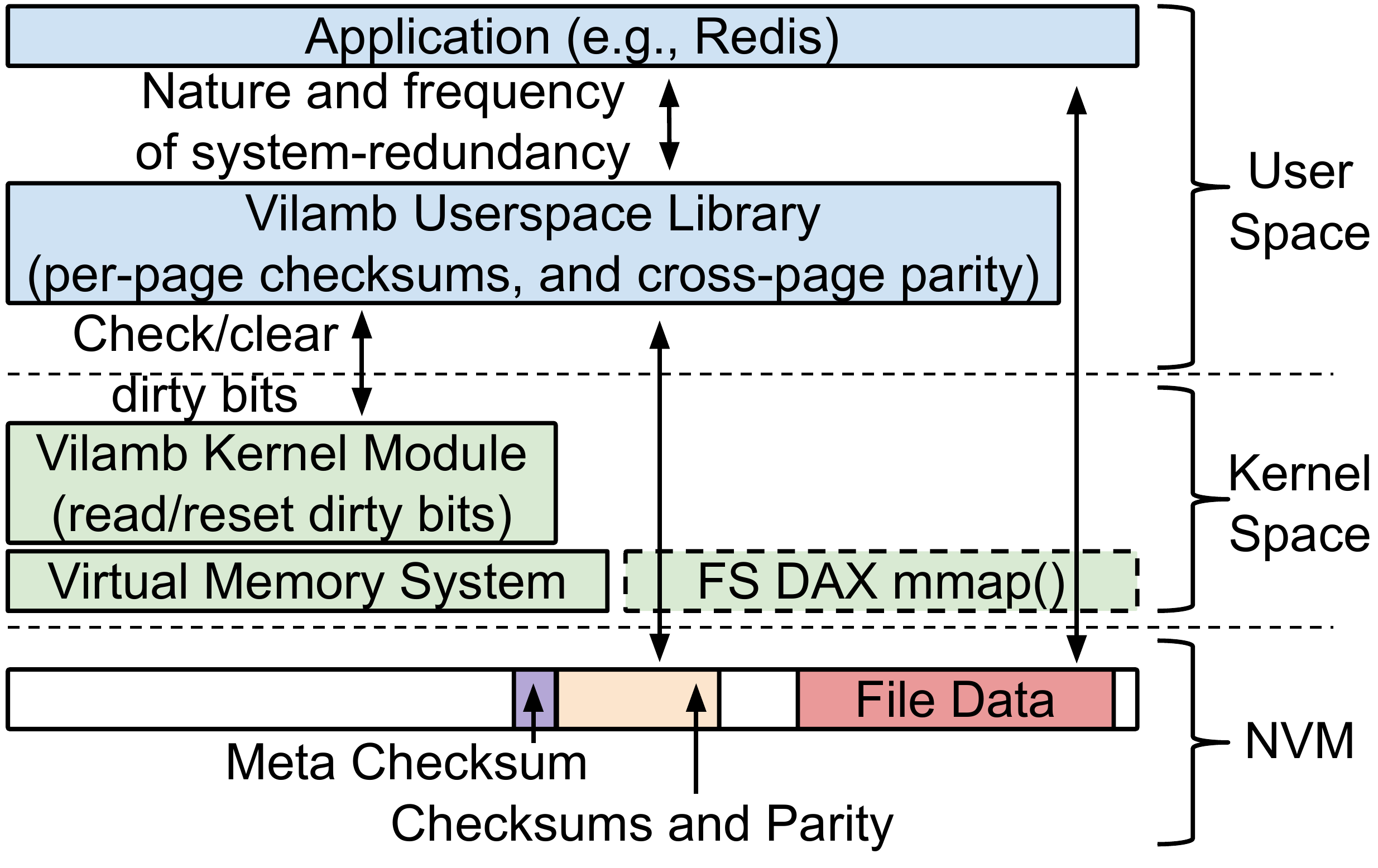}
\caption{\textbf{\system's Implementation}: The user space library 
performs the checksum and parity computations with a period that is 
set by the application. 
The kernel module checks and clears the dirty 
bits when requested by the user space library. }
\vspace*{-5mm}
\label{fig:impl}
\end{figure}

\subsection{Implementation}
\label{sec:implementation}
We implement \system as a user-space library. 
The library exposes an API 
that applications can use to configure the 
nature of \redundancy (e.g., type of checksum and 
number of pages in a stripe) and its update frequency. 
The library uses a periodic background thread that 
checks and clears 
the dirty bits using new system calls that we implement, and 
performs the \redundancy updates for the dirty pages. 
Our implementation uses a stripe size of five pages by default, 
with four consecutive data pages and one parity page. The stripes 
are statically determined at the time of initialization.
\cref{fig:impl} shows the components of our implementation. 

\newcommand{\getdb}{\texttt{getDirtyBits}\xspace}
\newcommand{\cleardb}{\texttt{clearDirtyBits}\xspace}
\textbf{New System Calls}: We implement two new system calls, 
\texttt{getDirtyBits} and \texttt{clearDirtyBits}, 
to check and clear the dirty bits for pages in a memory range, 
respectively. 
\getdb returns a bitvector that has
the dirty bits for pages in the input memory range. 
\cleardb accepts a dirty bitvector as its parameter in addition to 
a memory range. It clears the dirty bit 
for a page 
in the memory range only if the corresponding 
bit is set in the input dirty bitvector. 
Since \system is unaware of pages dirtied in between the checking and 
clearing and will not update their \redundancy, it uses 
this input dirty bitvector for \cleardb to 
clear the dirty bits only for pages that were  
dirty when initially checked. 

\textbf{Batched Checking and Clearing}:
\system checks and clears dirty bits
for multiple NVM pages (e.g., 512 in our experiments) as a 
batch for efficiency.
Both checking and clearing of dirty bits require a system call 
and traversing the hierarchical page table; clearing 
dirty bits further requires invalidating the corresponding TLB 
entries. Each of these is a costly operation, 
as evinced by prior research~\cite{optimizing-tlb-shootdown-atc}, 
and demonstrated by our experiments (\cref{sec:breakdown}).  
Batching allows pages to share 
the system call, fractions of the page table walk, and the 
TLB invalidation. We found that batching 
reduced the amount of time spent in checking/clearing 
dirty bits by up to two orders of magnitude.

\textbf{Algorithm}: \cref{impl:csum-parity-thread} 
details the steps that \system's  
background thread performs on each invocation. 
\system loops over all the $N$ pages in a given
DAX NVM file in increments of $B$ pages; 
$B$ being the batch size for which \system checks 
the dirty bits using a single 
system call (\cref{impl:get-dirty-bits}). 
\system stores a persistent shadow copy of the dirty bits 
(\cref{impl:copy-dirty-bits}) and then 
clears them (\cref{impl:clear-dirty-bits}). 
\system updates the checksum of each dirty 
page (\cref{impl:compute-csum}), and the parity 
of a group of $P$ pages if either of them is dirty 
(\cref{impl:compute-parity}). \system stores the 
checksums and parity separately from the 
data (\cref{fig:impl}) and then 
clears the shadow copy of the dirty bits 
(\cref{impl:clear-copy-dirty-bits}). 
\system then updates a meta-checksum (checksum of the page 
checksums) 
after every iteration
(\cref{impl:meta-checksum} and \cref{fig:impl}).

\setlength{\textfloatsep}{0pt}
\begin{algorithm}[t]
	\footnotesize
	\newcommand\mycommfont[1]{\footnotesize\textcolor{blue}{#1}}
	\SetCommentSty{mycommfont}
	\SetKwInOut{Parameter}{Parameter}
	\SetKw{KwBy}{increment by}

	\Parameter{Batch Size, B}
	\Parameter{Number of Pages in File, N}
	\Parameter{Number of Pages in a Parity Group, P}

	\SetKwData{DirtyBV}{dirtyBitvector}
	\SetKwData{CopyBV}{dirtyBitvectorCopy}
	\SetKwData{UpdateParity}{updateParity}
	\SetKwData{BatchStartPage}{currentBatchStartingPage}

	\SetKwFunction{GetDirtyBits}{checkDirtyBits}
	\SetKwFunction{ClearDirtyBits}{clearDirtyBits}
	\SetKwFunction{BitIsSet}{bitIsSet}
	\SetKwFunction{ComputeChecksum}{computePageChecksum}
	\SetKwFunction{ComputeMetaChecksum}{computeMetaChecksum}
	\SetKwFunction{ComputeParity}{computeParity}

	\For{$i\leftarrow 0$ \KwTo $N$ \KwBy $B$}
	{
		\DirtyBV $\leftarrow$ \GetDirtyBits{$i$, $i+B$}\; \label{impl:get-dirty-bits}
		\CopyBV $\leftarrow$ \DirtyBV\;\label{impl:copy-dirty-bits}
		\BatchStartPage $\leftarrow$ $i$\;\label{impl:start-batch-page}
		memoryFence\;\label{impl:memfence}
		\ClearDirtyBits{$i$, $i+B$, \DirtyBV}\;\label{impl:clear-dirty-bits}
		\For{$j\leftarrow i$ \KwTo $i+B$ \KwBy $P$}
		{
			\For{$k\leftarrow j$ \KwTo $j+P$ \KwBy $1$}
			{
				\UpdateParity $\leftarrow$ False\;
				\If {\BitIsSet{\DirtyBV, $k-i$}}
				{
					\UpdateParity $\leftarrow$ True\;
					\ComputeChecksum{$k$}\;\label{impl:compute-csum}
				}
			}
			\If {\UpdateParity}
			{
				\ComputeParity{$j$, $j+P$}\;\label{impl:compute-parity}
			}
		}
		memoryFence\;\label{impl:fence}
		\CopyBV $\leftarrow$ 0\;\label{impl:clear-copy-dirty-bits}
	}
	\ComputeMetaChecksum{}\;\label{impl:meta-checksum}
	\caption{System-Redundancy Update Thread}
	\label{impl:csum-parity-thread}
\end{algorithm}

As a performance optimization, instead of storing a shadow 
copy of the dirty bit for each page, we use 
a single dirty bitvector of size $B$ 
along with the current batch's starting page number 
(\cref{impl:copy-dirty-bits} and~\cref{impl:start-batch-page}). 
Together, the starting page number and the dirty bitvector copy 
suffice to store shadow copies of the dirty bits 
for pages in the current batch; pages not in the current batch 
do not need a shadow copy of their dirty bits because their dirty 
bits are not being cleared. Having a single dirty bitvector improves 
performance by reducing
cache pollution.

\system's redundancy verification thread (i.e., the scrubbing thread) 
computes and verifies the checksum only for pages that are clean, 
i.e., they have neither their dirty bit nor their shadown 
dirty bit set. If the checksum verification succeeds, the thread 
moves to the next page. In case of a checksum
mismatch, the scrubbing thread re-checks whether the page is 
clean. This second check is to ensure that 
the page was not modified after the first check but before 
the checksum verification. If the second check also 
indicates that the page is clean, 
the scrubbing thread raises a signal to halt the application. 
The file system can then recover the page, if it belongs 
to a clean stripe (we have not implemented recovery). 

\textbf{Leveraging Hardware Support}: Our implementation of \system leverages hardware-support 
whenever possible. We use
CRC-32C checksums and employ the \texttt{crc32q} instruction when available. 
Similarly, we use SIMD instructions for computing the parity 
whenever possible (e.g., by operating on 256-byte words in our 
experiments).
We never flush cache lines for persistence because we 
assume battery-backed servers. We do, however, use 
fences to ensure ordering between updates. For example, 
the fence at \cref{impl:memfence} ensures that the shadow 
copy of the dirty bits and current batch's starting page number 
writes are completed before the dirty bits are cleared. Similarly,
the fence at \cref{impl:fence} ensures that \redundancy is
written before the dirty bits' shadow copy is cleared.
We extend the same performance benefits 
(e.g., no cache line flushes and SIMD parity computations) to 
the alternatives that we compare \system with in our evaluation.

%% file: eval.tex
\section{Evaluation}
\label{sec:eval}
This section evaluates \system and 
compares it to \baseline and \sync, 
using eight macro- and micro-benchmarks.
\baseline serves as the baseline, providing 
the best performance but not implementing 
any \redundancy. 
Pangolin is a state-of-the-art userspace library that 
updates \redundancy when applications 
commit their data writes to NVM.

We obtained \sync's code from the authors 
and run it with checksum and parity 
updates enabled but checksum verification 
disabled (referred to as \sync-MLPC in 
the Pangolin paper~\cite{pangolin-atc}). 
We run \system also with checksum and parity updates 
enabled and checksum verification disabled. 
As shown in the evaluation of Pangolin~\cite{pangolin-atc}, 
and confirmed by our experiments, checksum verification 
via scrubbing at reasonable frequencies incurs 
negligible overhead. \sync can also verify 
checksums on object reads, which \system cannot, 
but doing so reduces throughput by up
to 50\% for large objects~\cite{pangolin-atc}.

Unless mentioned otherwise, \system
uses a 512-page batch size for checking/clearing dirty bits.
To accurately quantify \system's overheads, we pin 
it to the same core(s) as the application. For single 
threaded applications such as Redis, this means 
that the application and \system 
run on the same logical core (i.e., same hyper-thread). 
Each data point in our results 
is an average of three runs with root mean 
square error bars. 
We use a dual-socket 
Intel Xeon Silver 4114 machine with  
Linux 4.4.0 kernel for our experiments. The system has 192~GB DRAM, 
from which we use 64~GB as emulated NVM~\cite{pmem-emulation}.

\begin{figure*}[t]
	\framebox[\textwidth][c]{
	\includegraphics[width=0.24\textwidth, keepaspectratio]{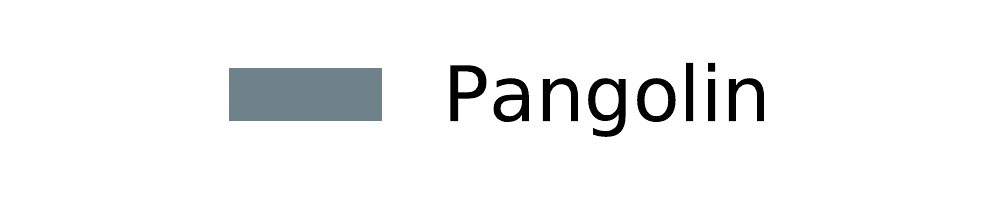}
	\includegraphics[width=0.5\textwidth, keepaspectratio]{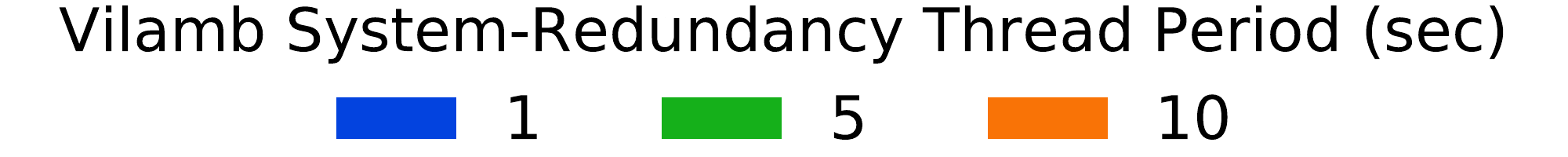}
	\includegraphics[width=0.24\textwidth, keepaspectratio]{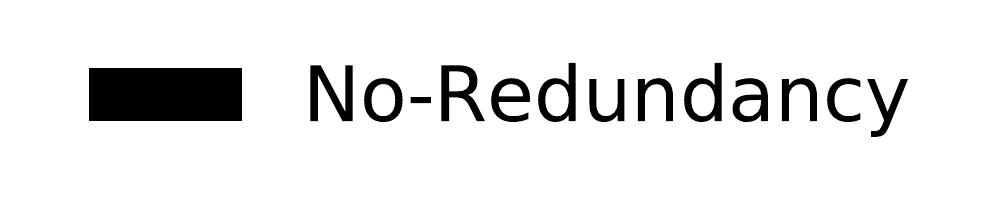}
	}
	\\[-1ex]
	 \begin{subfigmatrix}{3}
		\subfigure[Throughput]{\label{fig:redis-tput}
			\includegraphics[height=\linewidth, keepaspectratio]{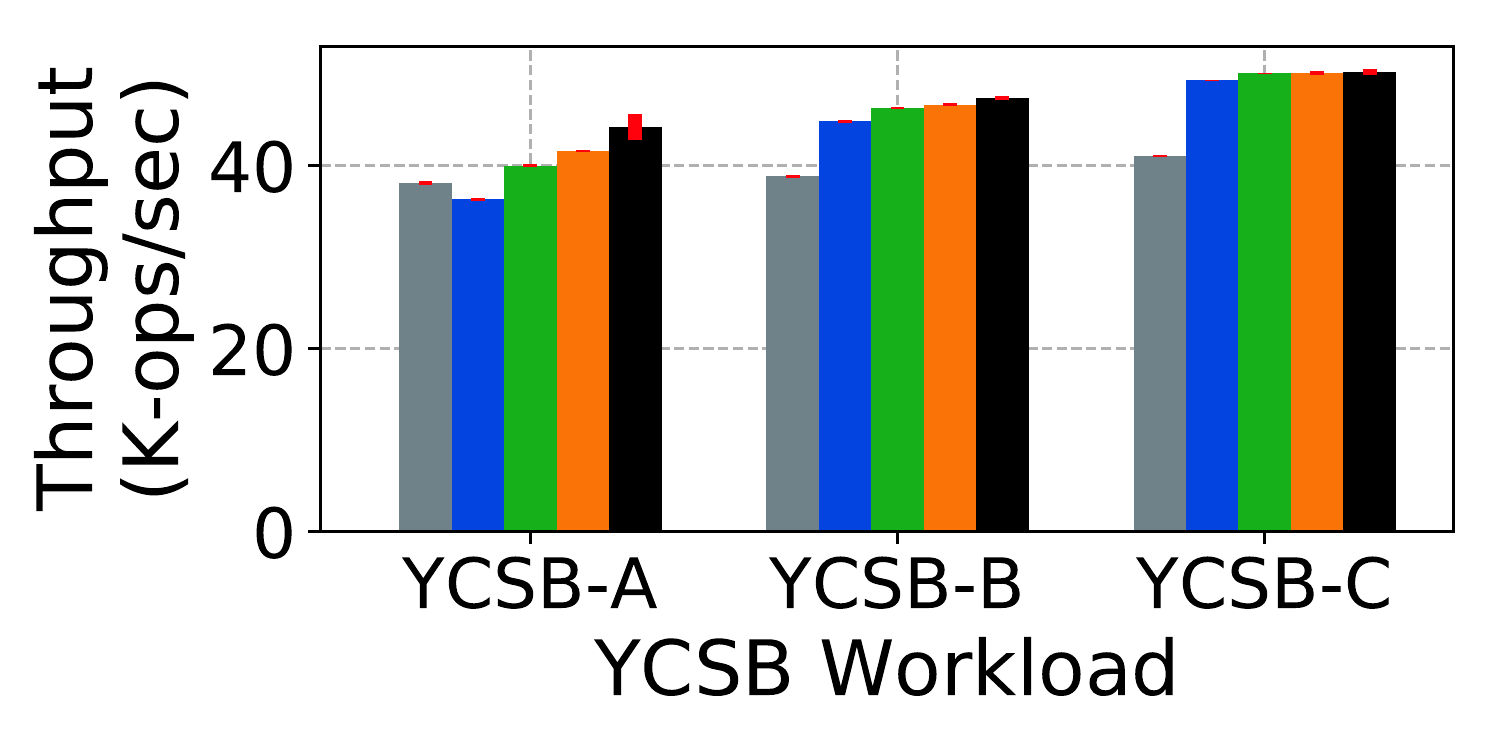}}
		\subfigure[Average Latency]{\label{fig:redis-avg-latency}
			\includegraphics[height=\linewidth, keepaspectratio]{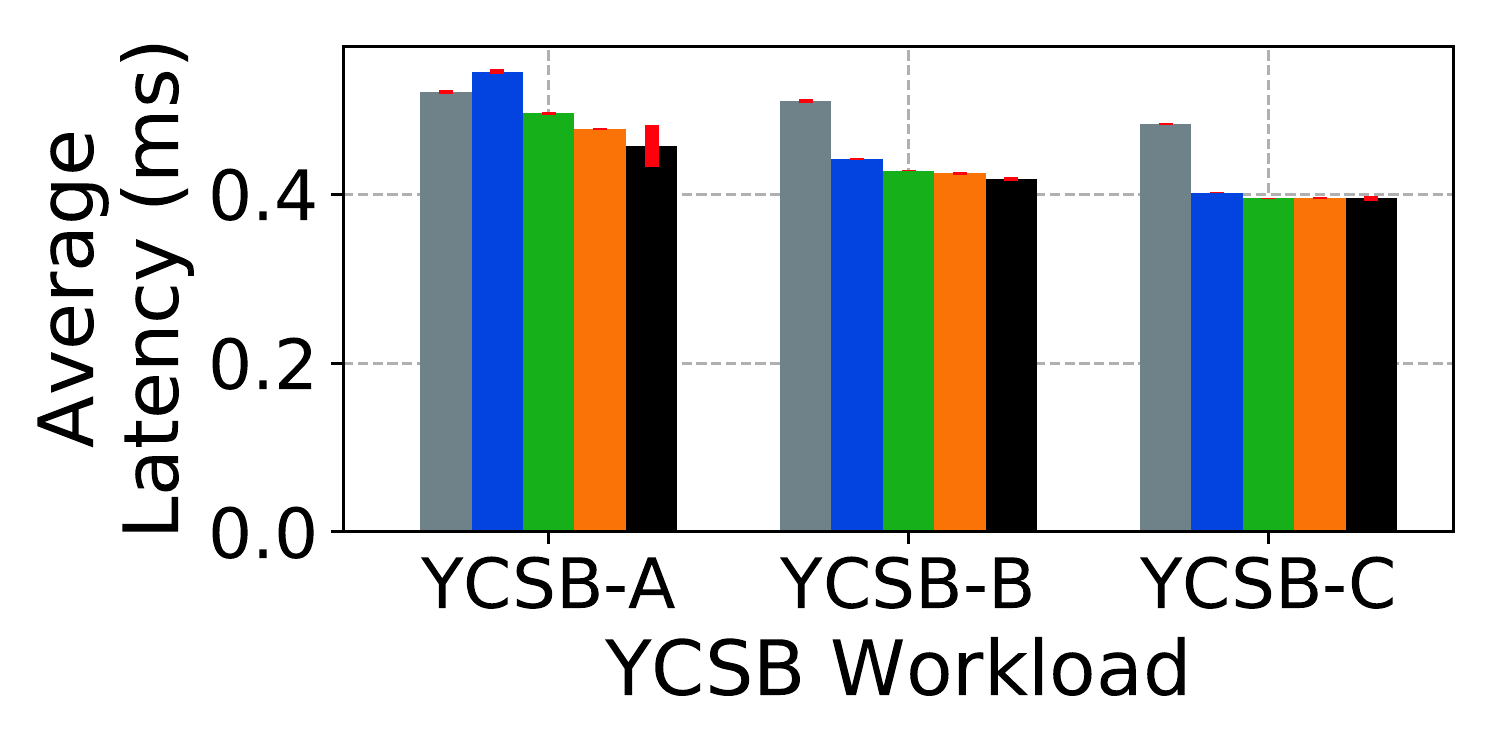}}
		\subfigure[Tail Latency]{\label{fig:redis-tail-latency}
			\includegraphics[height=\linewidth, keepaspectratio]{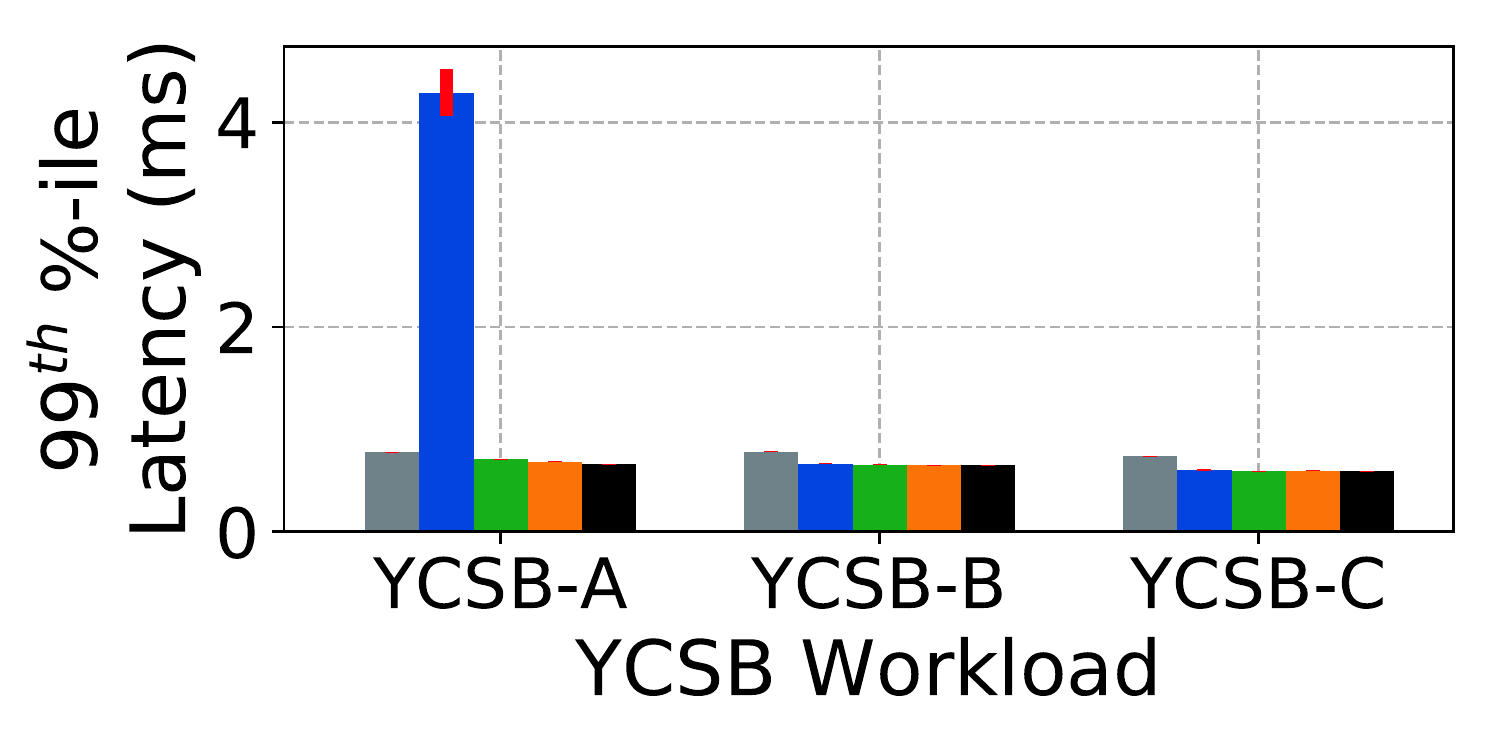}}
	 \end{subfigmatrix}
\caption{\textbf{YCSB with Redis} -- Throughput 
and read latency of YCSB workloads with Redis.}
\label{fig:redis-zipf}
\end{figure*}

\subsection{Key Evaluation Takeaways}
\label{sec:eval-takeaways}
Key takeaways from 
our evaluation include: 

\setlist{nolistsep}
\begin{itemize}
	\item \system is low-overhead. 
For example, \system with a 10~sec \redundancy 
update period reduces Redis' YCSB throughput by only 0.1--6\%
in comparison to \baseline. 
\item \system significantly outperforms 
\sync. For example, \system has 3--5$\times$ higher 
insert throughput than \sync for five PMDK key-value stores.
Even for low throughput applications like 
single threaded Redis serving YCSB, \system has up to 18\% 
higher throughput than \sync.
\item \system significantly increases the MTTDL. For example, 
	\system increases the MTTDL for PMDK key-value stores by 
	up to two orders of magnitude. 
\item \system offers a tradeoff between
		performance and time-to-coverage. 
		For example, decreasing  
		the delay between \redundancy updates 
		from 5~sec to 1~sec increases Redis' YCSB-A MTTDL by 3$\times$ but 
		decreases the throughput by 10\%.
	\item \system's battery requirements are low. Across 
		all of our workloads, the cost of batteries 
		that \system requires never exceeds \$10.
\end{itemize}

\subsection{YCSB with Redis}
\label{sec:macrobench}
Redis~\cite{Redis} is a 
widely used 
open-source NoSQL DBMS. We modify it
to use a DAX NVM file for its data heap.
Our implementation  
uses the \texttt{libpmemobj} library~\cite{libpmemobj} from the 
Intel persistent memory development kit (PMDK)~\cite{pmemlib} 
for \baseline. 

\textbf{Modifying Redis to use \system and \sync}:
For \system, we added 10 lines of 
initialization and cleanup code 
in one file. The initialization code registers 
Redis' NVM heap with \system and sets the 
\redundancy update delay. 
To use \sync's 
transactional API (which is similar to but 
different than \texttt{libpmemobj}), we changed 
346 lines of code across 10 files 
in Redis. Whereas most of these changes were to 
the transactional interface (e.g., using \texttt{pgl\_tx\_begin}),
we also had to modify Redis to invoke \sync
before reading data from an object (using \texttt{pgl\_get}). 
Doing so enables \sync to determine whether 
the object is in NVM or in DRAM and provide Redis 
with the correct pointer. 

\textbf{Experimental Setup}:
We use three core YCSB workloads:
YCSB-A (50:50 reads:updates), 
YCSB-B (95:5 reads:updates), and YCSB-C (read-only).
We initialize the DBMS with 
1M ($1\times2^{20}$) key-value pairs 
for a NVM footprint of 10~GB
and run the workloads for 
five minutes. 
The YCSB workload generator uses 20 threads 
and runs on a different socket than Redis.

\begin{figure*}[t]
	\framebox[\textwidth][c]{
	\includegraphics[width=0.24\textwidth, keepaspectratio]{./graphs/legend_pgl_redis.pdf}
	\includegraphics[width=0.5\textwidth, keepaspectratio]{./graphs/legend_vilamb_redis.pdf}
	\includegraphics[width=0.24\textwidth, keepaspectratio]{./graphs/legend_pmemobj_redis.pdf}
	}
	\\[-1ex]
	 \begin{subfigmatrix}{2}
		\subfigure[Insert Throughput]{\label{fig:insert-kvs}
			\includegraphics[height=\linewidth, keepaspectratio]{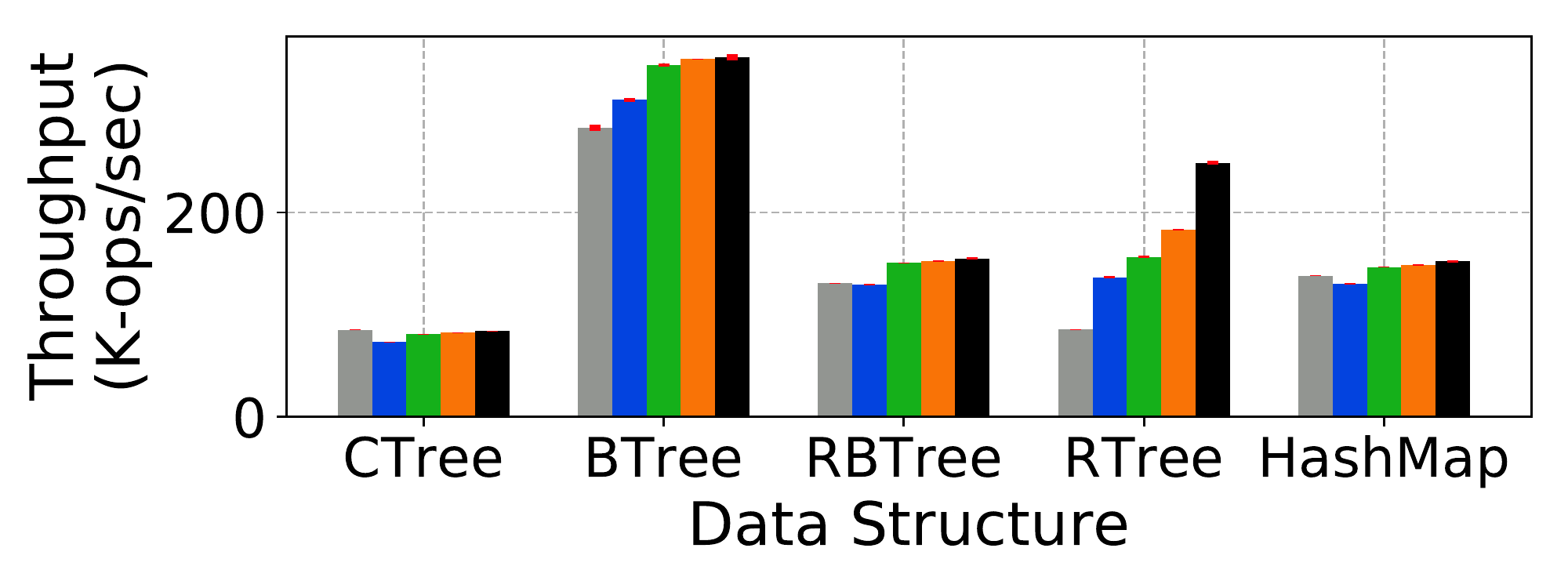}}
		\subfigure[Remove Throughput]{\label{fig:remove-kvs}
			\includegraphics[width=0.45\textwidth, keepaspectratio]{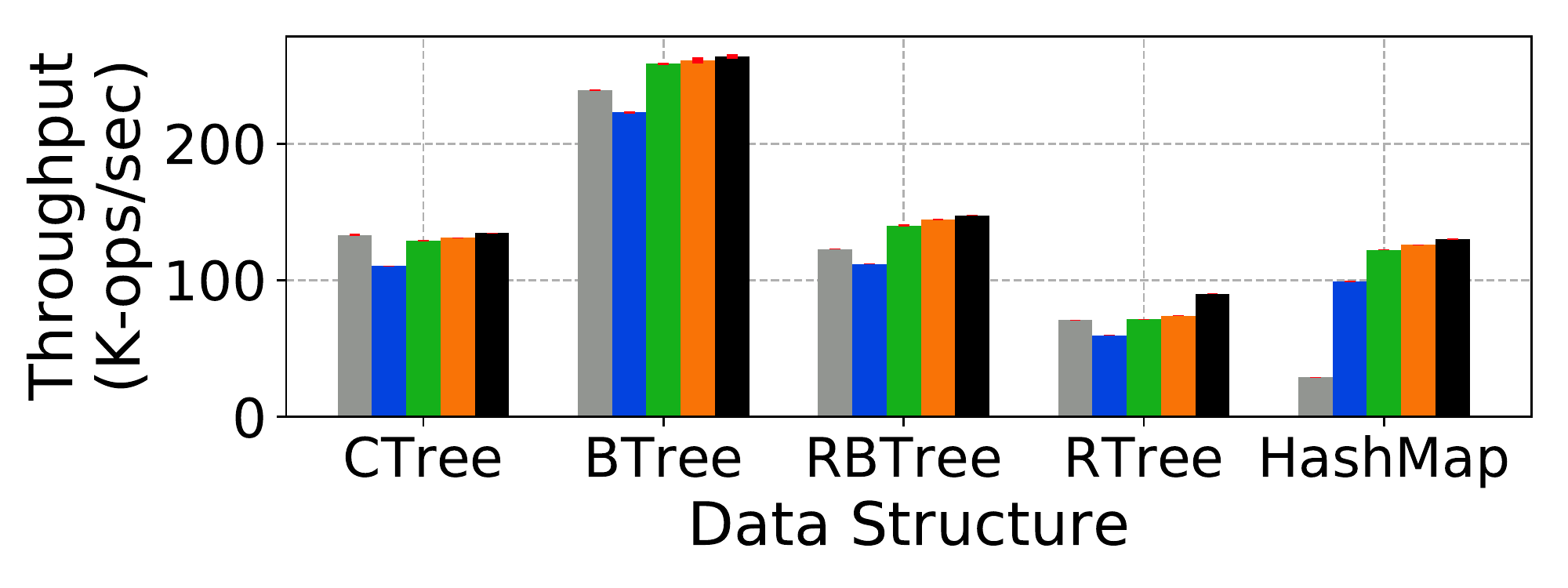}}
	 \end{subfigmatrix}
	 \center{\framebox[0.7\textwidth][c]{
		\includegraphics[width=0.7\textwidth, keepaspectratio]{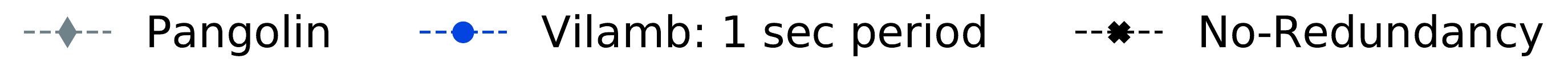}
	  }}
	\\
	 \begin{subfigmatrix}{5}
		\subfigure[CTree Insert]{\label{fig:ctree-tput}
			\includegraphics[height=\linewidth, keepaspectratio]{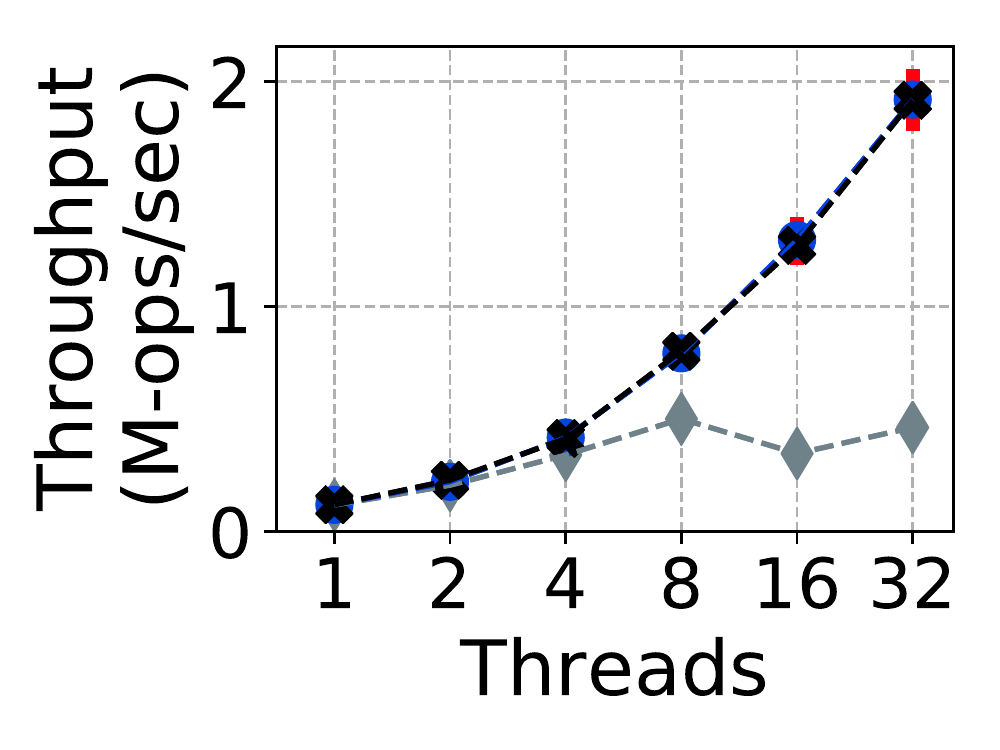}}
		\subfigure[BTree Insert]{\label{fig:btree-tput}
			\includegraphics[height=\linewidth, keepaspectratio]{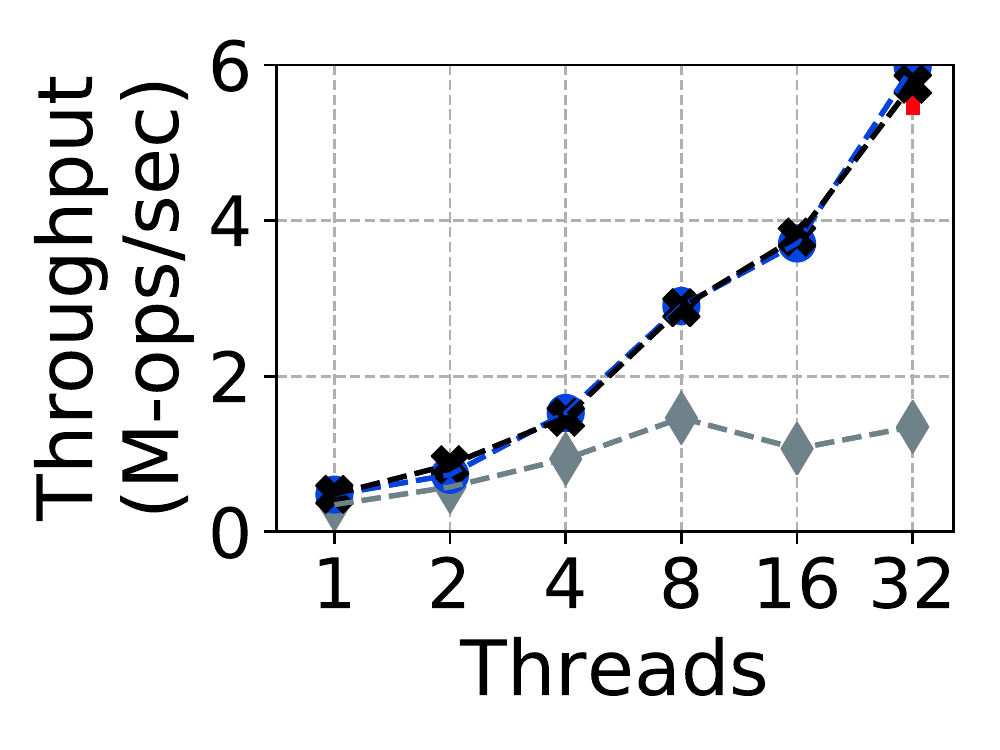}}
		\subfigure[RBTree Insert]{\label{fig:rbtree-tput}
			\includegraphics[height=\linewidth, keepaspectratio]{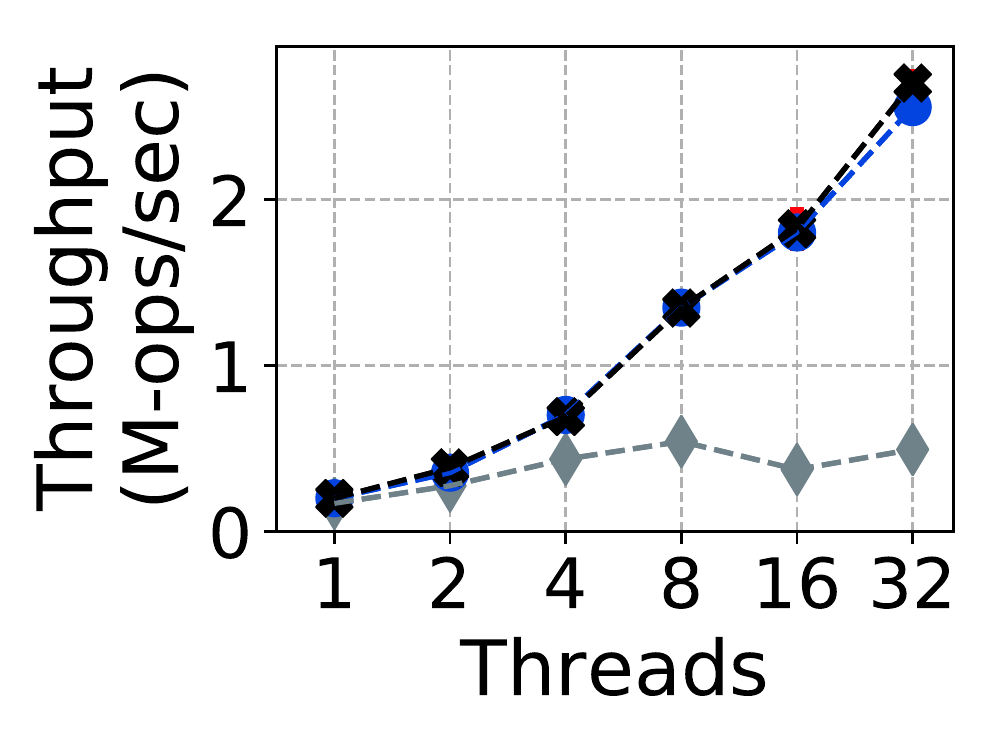}}
		\subfigure[RTree Insert]{\label{fig:rtree-tput}
			\includegraphics[height=\linewidth, keepaspectratio]{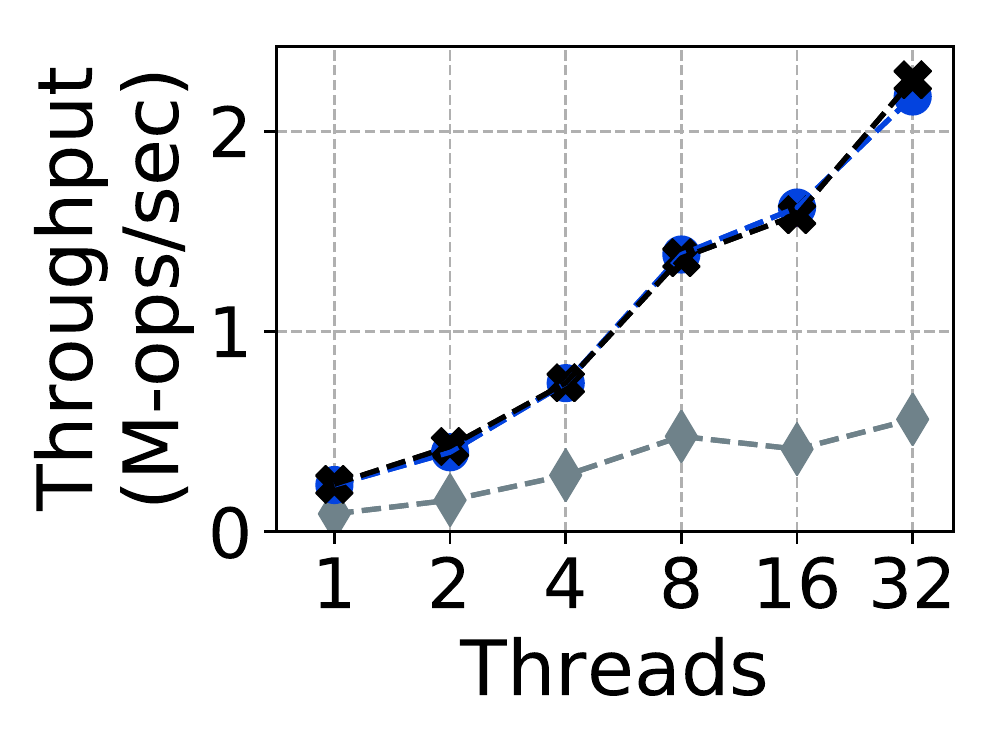}}
		\subfigure[HashMap Insert]{\label{fig:hashmap-tput}
			\includegraphics[height=\linewidth, keepaspectratio]{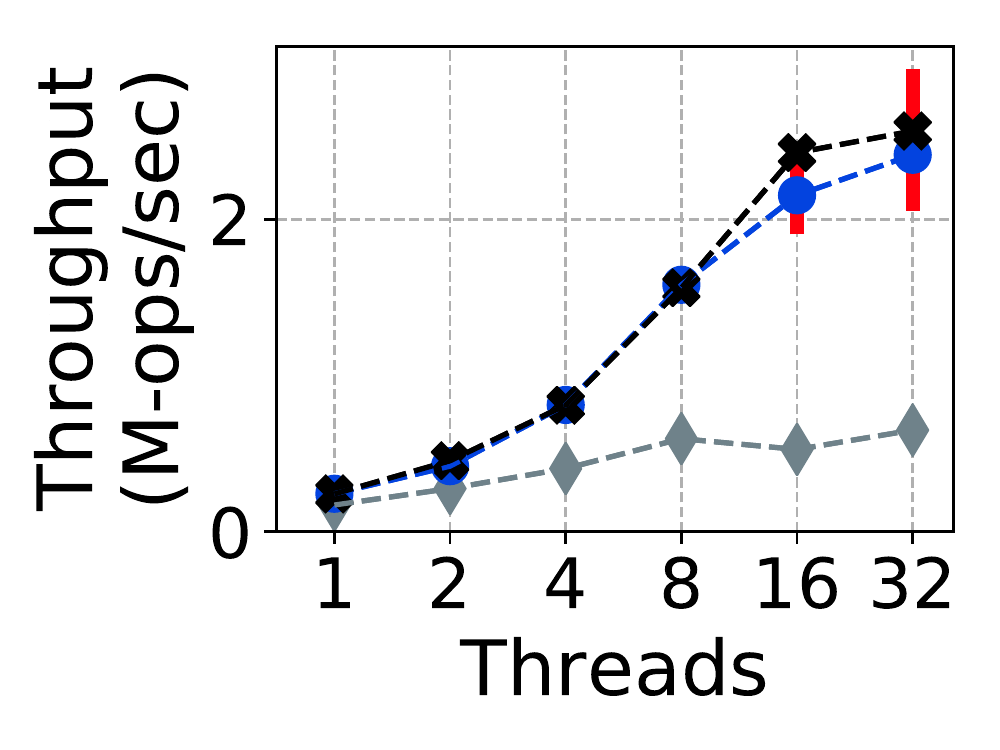}}
	 \end{subfigmatrix}
\caption{\textbf{PMDK Key-Value Stores} -- Throughput for 
	insert-only, remove-only benchmarks with different 
	PMDK key-value stores.}
\label{fig:map-bench}
\end{figure*}

\textbf{Results}: \cref{fig:redis-zipf} presents 
throughput and read latencies.
\system reduces the throughput, in comparison to 
\baseline, by 0.1--6\% for a 
\redundancy update period of 10~sec and by 1.6--17\%
for a period of 1~sec. 
Increasing the delay for \redundancy updates 
improves \system's performance because 
it performs fewer 
\redundancy updates and 
hogs less CPU.
With aggressive \redundancy updates every 
second, \system increases the tail 
latency for YCSB-A because it stalls 
Redis while updating \redundancy 
on the same core. 
This effect can be mitigated 
if \system and Redis were to run on separate cores.

\sync's throughput is 13--18\% lower than \baseline, with 
a higher overhead for more read-heavy workloads.
In addition to the overhead of updating \redundancy, 
\sync incurs overhead because of two other factors, 
both related to its micro-buffering 
design. First, on every object read, 
\sync probes a cuckoo hash table to check 
whether the latest copy of the object 
is in a DRAM micro-buffer or in NVM. Second, 
when Redis adds an object to a transaction,
\sync copies the entire object to DRAM for 
micro-buffering, rather than just the 
modified data ranges. 

For the write-heavy workload YCSB-A, \sync outperforms 
\system with a \redundancy update period of 1~sec. 
This is because \sync's micro-buffering design 
enables it to perform checksum and parity 
updates using the diff of the updated data. 
\sync uses the new data in the DRAM micro-buffer and 
the old data in the NVM to compute the data diff.
In contrast, \system has 
to read the entire page to update the checksum, and 
also read other pages in the stripe to update the parity.
With 5 and 10~sec \redundancy update periods, 
\system outperforms \sync by 5--7\%.

For read-heavy workloads YCSB-B and YCSB-C, 
\system reduces the throughput marginally (e.g., 
less than 2\% for YCSB-C) whereas \sync reduces the throughput 
by 18\%. This is because even though the number of 
\redundancy updates reduce, \sync continues to 
incur the additional overheads described above. 
For example, \sync has to check whether the data is in 
DRAM or NVM for object reads. 


\sync's moderate overhead (up to 18\%) 
compared to \baseline and \system
is an artifact of Redis' inefficiencies. In 
particular, Redis' single-threaded design 
causes it to have low performance 
(tens of thousands of operations per sec) that 
does not fully expose the \redundancy update overheads.
In the next section, we show that 
multi-threaded key-value stores that perform 
millions of operations per second benefit 
significantly from \system's asynchronous approach.


\subsection{PMDK Key-Value Stores}
\label{sec:kvs}
Intel persistent memory development kit (PMDK)~\cite{pmemlib} 
implements NVM-optimized key-value stores and 
includes performance benchmarks. 

\textbf{Experimental Setup}:
Similar to Pangolin~\cite{pangolin-atc}, 
we use insert-only, and remove-only 
benchmarks for five key-value stores: 
Crit-Bit Tree (CTree), 
BTree, Red-Black Tree (RBTree), Range Tree (RTree) 
and chaining hashmap (HashMap). 
We first re-create the experiment 
and results from Pangolin~\cite{pangolin-atc} 
with a single-thread that performs 
5 million operations. 
We then use multiple threads (1 to 32) 
with 100,000 operations per 
thread. 

We modify the PMDK benchmark for multi-threaded 
benchmarking. In the original implementation, 
the threads synchronize using a coarse-grained 
lock; each thread holds a lock over the entire 
data structure for the entire duration of its 
transaction. Not surprisingly, the coarse-grained lock 
leads to poor scaling. We modified the implementation
such that each thread maintains and operates on its 
own instance of the data structure. All the threads 
share the same NVM pool, but do not synchronize 
their changes because they operate on different data. 
Our modifications enabled close to linear scaling 
for the baseline case of \baseline.

\textbf{Results}: 
\cref{fig:insert-kvs,fig:remove-kvs} show the throughput 
for the insert-only and remove-only workloads 
when using a single thread for the key-value store.
\sync's overheads are similar to those reported 
in their paper~\cite{pangolin-atc}. 
\system's performance improves 
with increasing delay in \redundancy updates. 
Of the five key-value stores, both \sync 
and \system have the highest overhead in comparison 
to \baseline for RTree because RTree's 
insertion touches the largest amount of data. 
For the remove-only workload, \sync outperforms 
\system with 1~sec \redundancy update period 
because removing objects touches only a small amount of 
data and \sync can efficiently update \redundancy using 
the diffs for small data. 

\cref{fig:ctree-tput,fig:btree-tput,fig:rbtree-tput,fig:rtree-tput,fig:hashmap-tput} 
show the insert-only 
throughput for the five key-value stores with increasing 
number of threads. Increasing the number of 
threads 
updates
NVM data more aggressively and generates more 
\redundancy updates. This 
causes \sync to have up to 80\% lower throughput than 
\baseline. 
Across the the five key-value store, \system 
has 3--5$\times$ higher throughput than 
\sync when using 32 threads. 

\begin{figure*}[t]
	\framebox[\textwidth][c]{
	\includegraphics[width=0.24\textwidth, keepaspectratio]{./graphs/legend_pgl_redis.pdf}
	\includegraphics[width=0.5\textwidth, keepaspectratio]{./graphs/legend_vilamb_redis.pdf}
	\includegraphics[width=0.24\textwidth, keepaspectratio]{./graphs/legend_pmemobj_redis.pdf}
	}
	 \begin{subfigmatrix}{3}
		\subfigure[Allocation]{\label{fig:alloc-latency}
			\includegraphics[height=\linewidth, keepaspectratio]{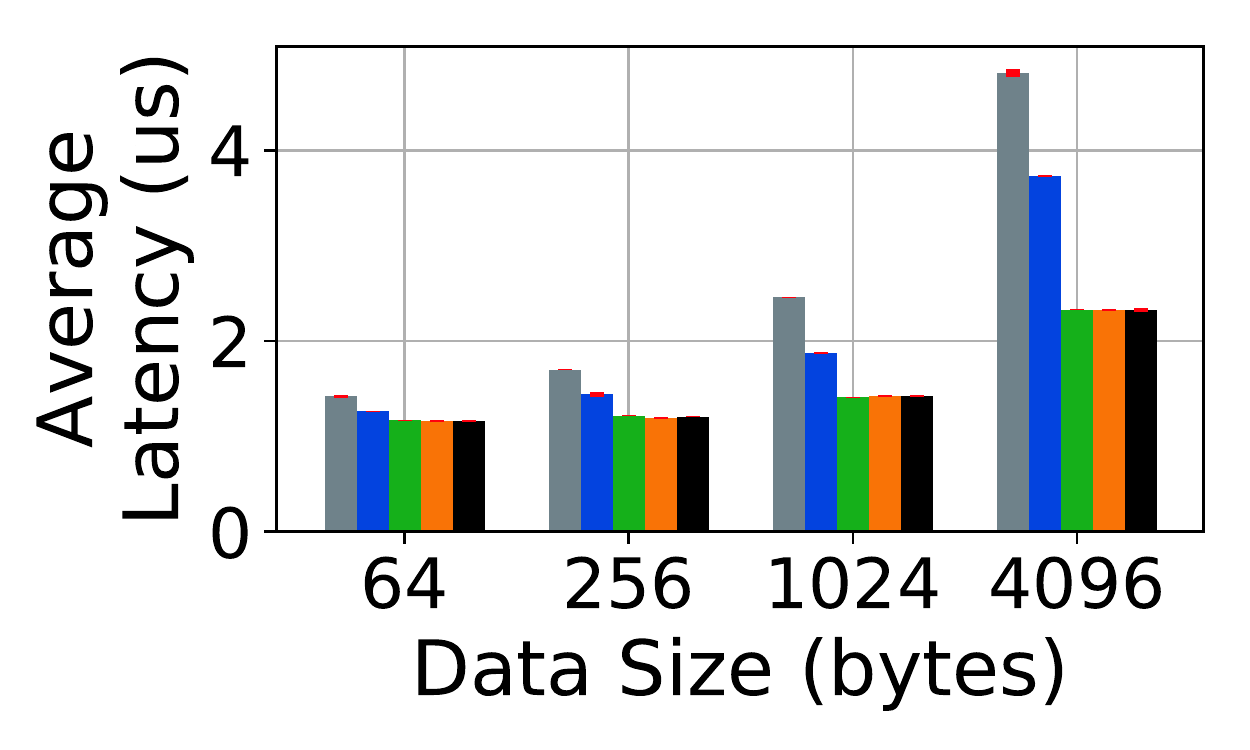}}
		\subfigure[Overwrite]{\label{fig:write-latency}
			\includegraphics[height=\linewidth, keepaspectratio]{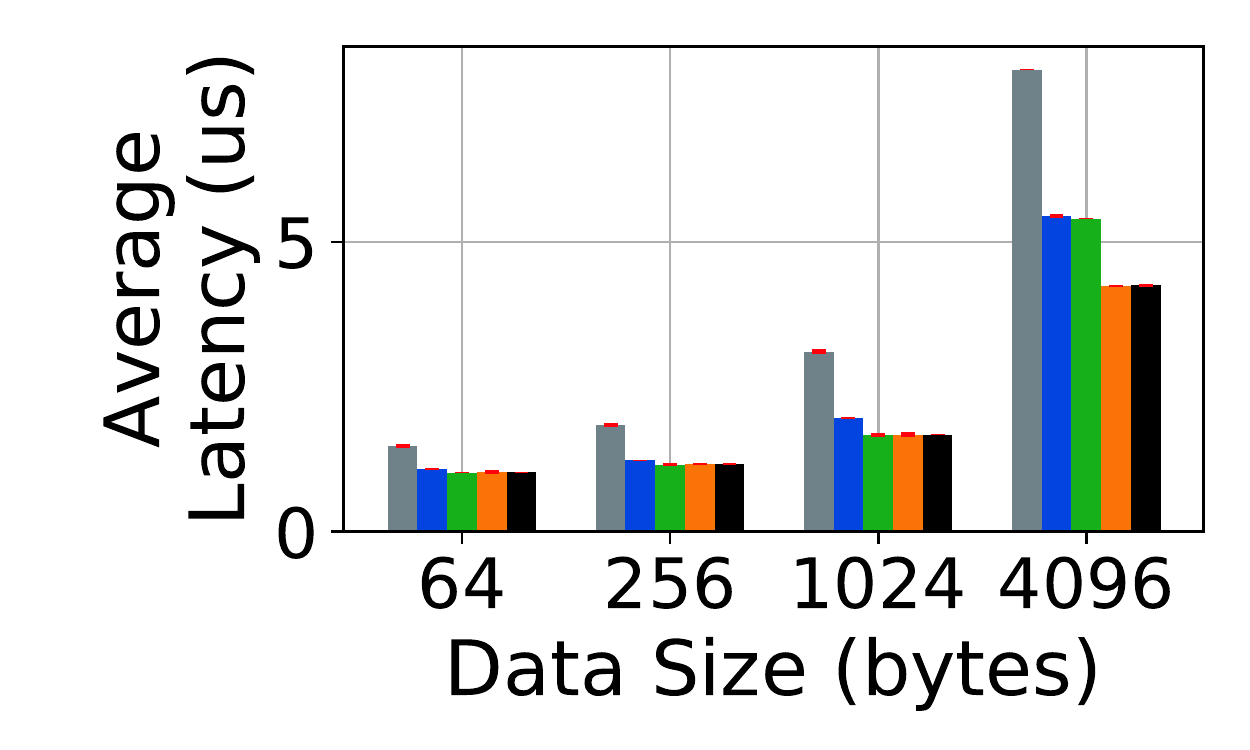}}
		\subfigure[Deallocation]{\label{fig:free-latency}
			\includegraphics[height=\linewidth, keepaspectratio]{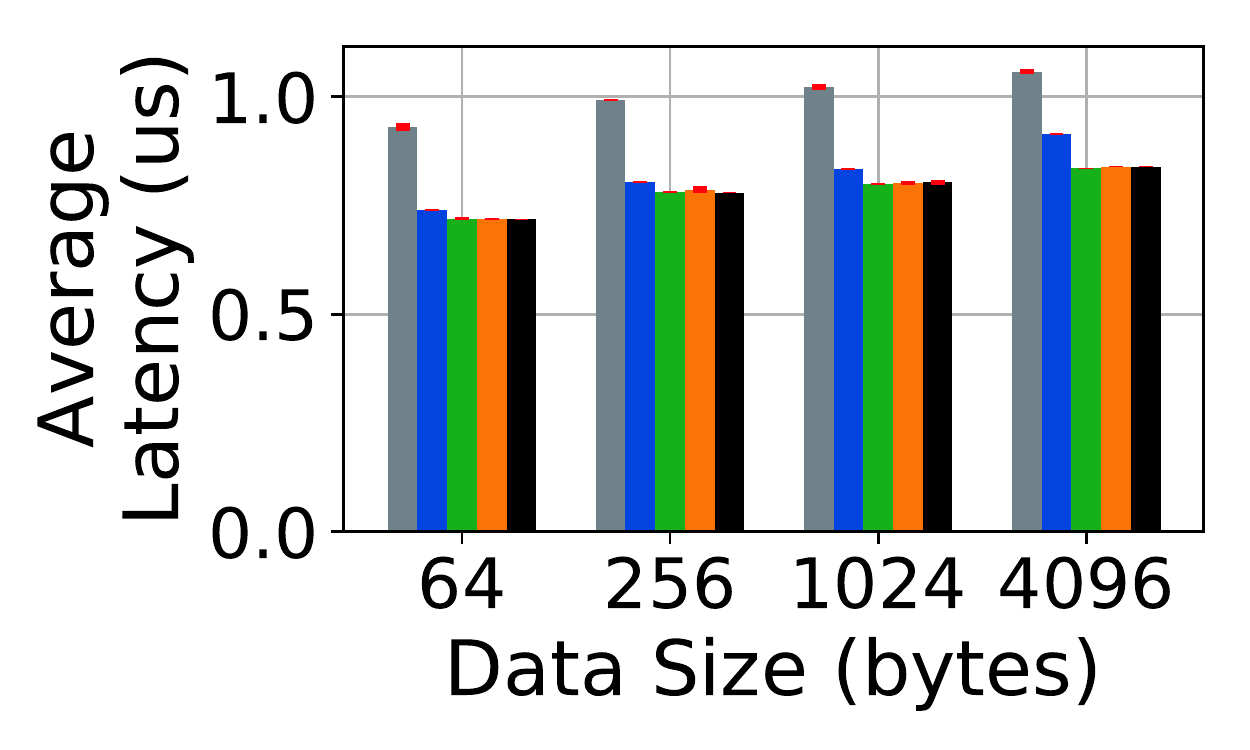}}
	 \end{subfigmatrix}
\caption{\textbf{NVM Transaction Latencies} -- Latencies  
	for transactional allocation, overwriting, and deallocation.}
\label{fig:nvm-tx-latencies}
\end{figure*}

\subsection{NVM Transaction Microbenchmarks}
\label{sec:op-bench}
Pangolin~\cite{pangolin-atc} introduced micro-benchmarks to 
measure the latency of transactional operations (allocation, 
overwrite, and deallocation), and to measure the scalability 
of overwriting NVM regions with multiple threads. 

\textbf{Experimental Setup}: We perform each 
transactional operation (allocation, overwrite, deallocation) 
1 million times for different sized objects in a single thread 
and report the average latency. We use an NVM file of
10~GB for this. For scalability, we increase the number 
of threads with each thread overwriting 64-byte 
and 4~KB regions 200,000 times.

\textbf{Results}:
\cref{fig:nvm-tx-latencies} shows the latency 
for performing the transactional operations 
using a single thread.
For 64-byte objects, \sync 
incurs 23\%, 44\%, and 30\%
higher latency than \baseline for allocation, overwrite, 
and deallocation, respectively. In contrast, \system 
with a \redundancy update period of 
1~sec increases the corresponding latencies by 
only 9\%, 5\%, and 3\%; increasing the 
\redundancy update period further reduces 
\system's latencies. 
Increasing the object sizes increases the latency 
for all configurations, because more data is 
touched (except for deallocation, in which only metadata 
is updated). However, even for 
4~KB objects, \system with a \redundancy 
update period of 1~sec has  
13\%--31\% lower latencies than 
than \sync. 

\begin{figure}[t]
	 \framebox[\columnwidth][c]{
		\includegraphics[width=\columnwidth, keepaspectratio]{./graphs/legend_map_threads.pdf}
	 }
	 \\[-1ex]
	 \begin{subfigmatrix}{2}
	    \subfigure[64 Byte Writes]{\label{fig:write-tput-64}
			\includegraphics[width=0.45\columnwidth]{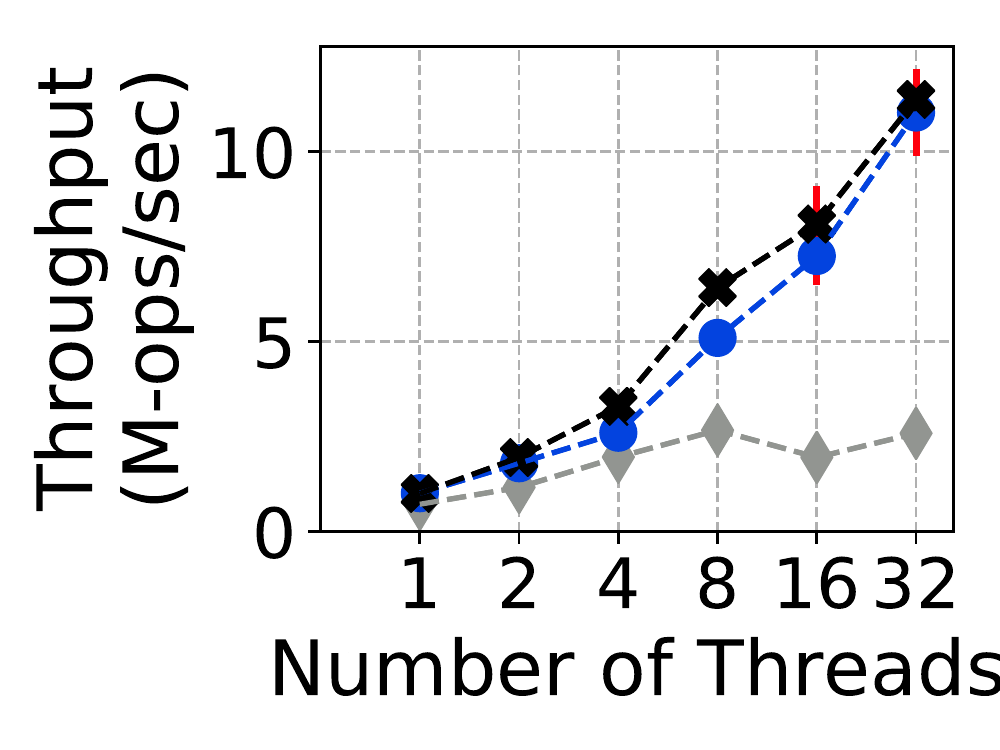}}
		\subfigure[4096 Byte Writes]{\label{fig:write-tput-4k}
			\includegraphics[width=0.45\columnwidth]{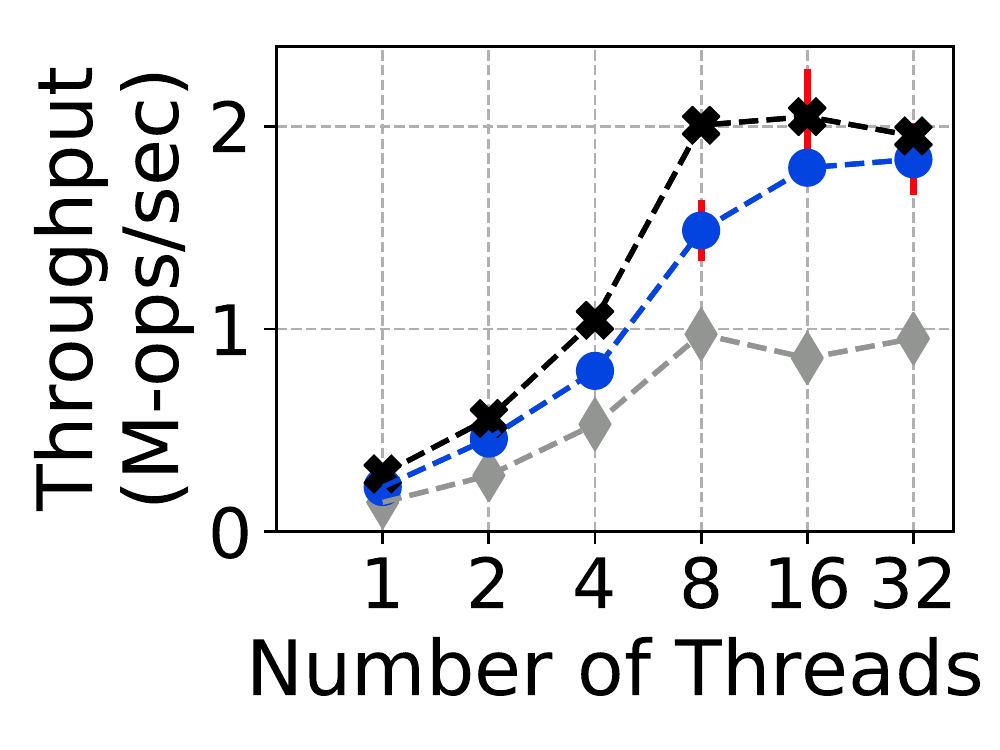}}
	 \end{subfigmatrix}
\caption{\textbf{NVM Overwrite Throughput}}
\label{fig:nvm-write-tput}
\end{figure}

\cref{fig:nvm-write-tput} shows the throughput 
for overwriting 64-byte and 4~KB regions with 
increasing number of threads. \system scales 
close to \baseline, with only up to 25\% lower 
throughput. In contrast, \sync has up to 77\% 
lower throughput. 
Pangolin's experiments 
with real NVM (in contrast to our 
DRAM-based emulation) showed that \baseline performance 
does not scale well beyond 8 threads because of 
NVM's limited bandwidth~\cite{pangolin-atc}.
However, even with 8 threads 
\system's throughput is double of \sync's. 
As NVM performance improves and gets closer 
to DRAM performance, the benefits of 
\system's asynchronous redundancy maintenance will become
more pronounced. We also evaluated overwriting 
with other intermediate data sizes (256 and 1024 bytes) and obtained 
similar trends.

\subsection{Fio Microbenchmarks}
\label{sec:fio}
This section evaluates \system's performance using \fio~\cite{fio} microbenchmarks.
We cannot evaluate \sync using \fio because \fio's NVM engine~\cite{fio-pmem} 
does not use object based transactions. Rather \fio treats the entire
DAX-mapped file as a raw sequence of bytes. This illustrates \sync's 
programming model restriction. Applications that manage DAX-mapped 
data themselves, either as raw data as in \fio microbenchmarks or 
in a more complex fashion like NVM databases~\cite{nstore-sigmod}, can 
benefit from \sync only if they can be and are
modified to use its APIs.

\textbf{Experimental Setup}:
Fio's libpmem engine reads/writes 
DAX NVM files at a cache line granularity. 
We use write-only and read-only workloads with a 16~GB file
and three access patterns: uniform random, sequential, and Zipf. 
The workloads perform reads/writes equal to the file size. 
The random and sequential workloads choose previously 
unread/unwritten cache lines, consequently reading/writing 
each cache line in the entire file exactly once. 
We use a single thread and pin it to a logical core along with \system.

\begin{figure}[t]
	\framebox[\columnwidth][c]{
	\includegraphics[width=0.65\columnwidth, keepaspectratio]{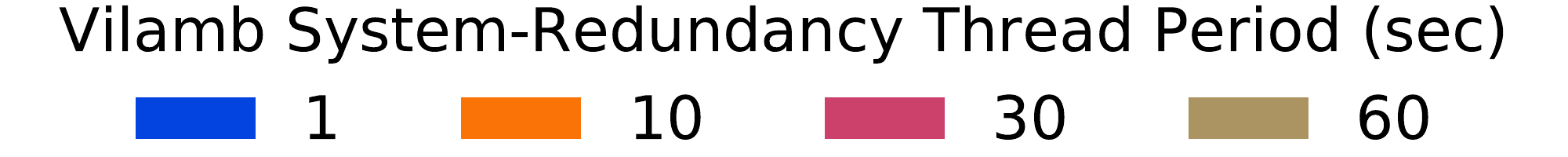}
	\includegraphics[width=0.35\columnwidth, keepaspectratio]{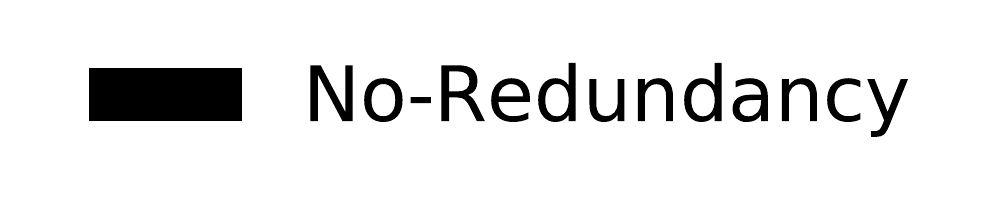}
	}
	\\[-1ex]
	 \begin{subfigmatrix}{2}
		\subfigure[Write Only Workload]{\label{fig:fio-write-only}
			\includegraphics[width=0.45\columnwidth, keepaspectratio]{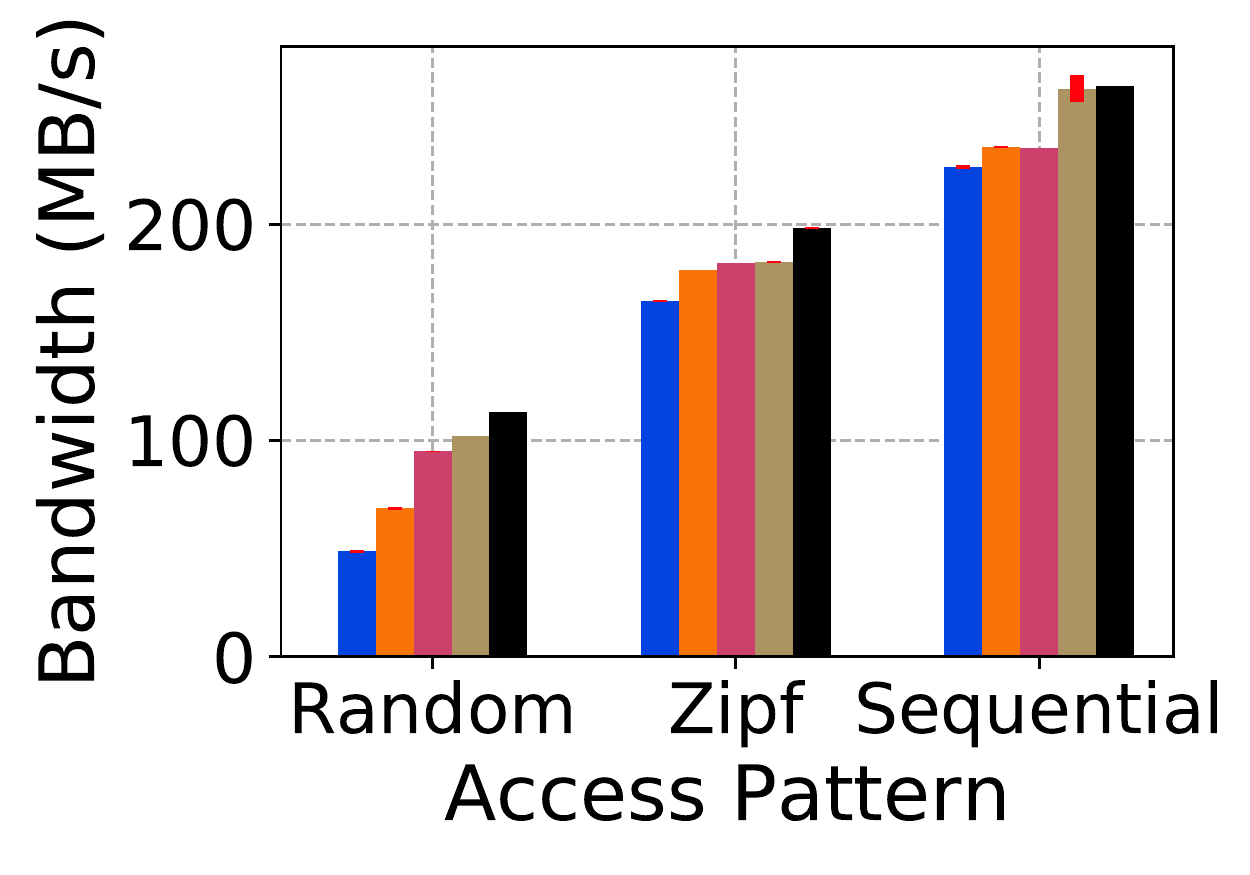}}
		\subfigure[Read Only Workload]{\label{fig:fio-read-only}
			\includegraphics[width=0.45\columnwidth, keepaspectratio]{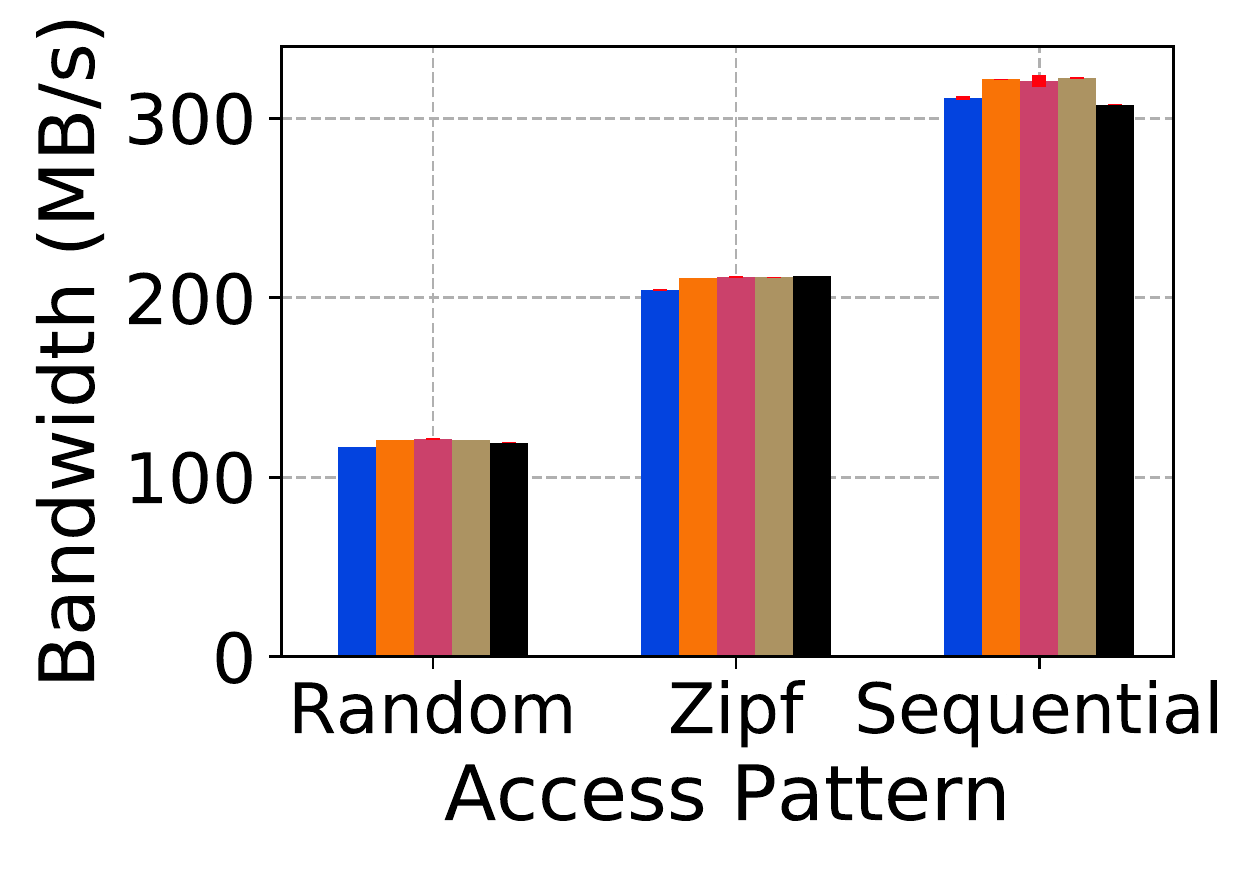}}
	 \end{subfigmatrix}
\caption{\textbf{Fio Microbenchmarks} -- Throughputs 
for write-only and read-only workloads with different access patterns.} 
\label{fig:fio}
\end{figure}

\textbf{Results}: \cref{fig:fio} shows the throughput 
for the two workloads with three access patterns each. 
For write-only workloads, \system reduces throughput by 
0.5--56\% with higher overheads for more frequent \redundancy updates.
\system's overheads are highest for the random workload and 
lowest for the sequential workload; sequential workloads offer the best 
opportunity to reduce computations, because successive cache line writes 
belong to the same page. Even for random workloads, 
the overhead is only 10\%
with a \redundancy update delay of  
60 seconds. 
\system reduces the throughput by only up to 3\%
for read-only workloads, demonstrating the 
efficacy of its checking of dirty bits.
\system's througput is higher than \baseline 
for the read-only sequential workload with an 
update period of more than 10 seconds; this is 
an artifact of the experimental setup.
While checking for dirty bits, \system populates the 
page table entries and reduces the number of soft 
page faults. The performance benefit of reduced 
soft page faults outweigh the overhead of checking 
the dirty bits infrequently (i.e., with a period of 
more than 10 seconds). This anamoulous 
inversion of performance can be resolved
by pre-populating the page table 
entries for \system as well.

\subsection{Cost of Checking/Clearing Dirty Bits}
\label{sec:breakdown}
To better understand the cost of checking and clearing dirty 
bits,
we break down the cost into its constituent components: 
(i) system call, (ii) page table
walk to desired page table entries, 
(iii) reading/resetting the dirty bits, and (iv)
TLB invalidation after clearing dirty bits.
We also demonstrate the benefits of batching multiple pages 
when checking and clearing the dirty bits. 

\textbf{Experimental Setup}: We use the write-only \fio workload 
with 64-byte writes and a uniform random access pattern. 
We configure \system to check/clear the dirty 
bits every second. We
measure the average amount of time spent in each of the 
components for a single invocation of \system's background thread. 
We vary the batch size 
to demonstrate the impact of batching. 

\begin{figure}[t]
	\makebox[\columnwidth][c]{
	\includegraphics[width=\columnwidth, keepaspectratio]{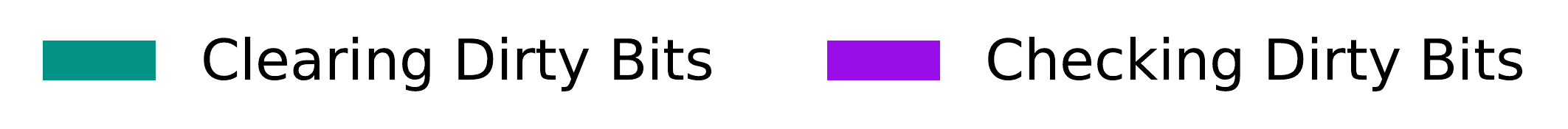}
	}
	\\[-1.5ex]
	\makebox[\columnwidth][c]{
	\includegraphics[width=\columnwidth, keepaspectratio]{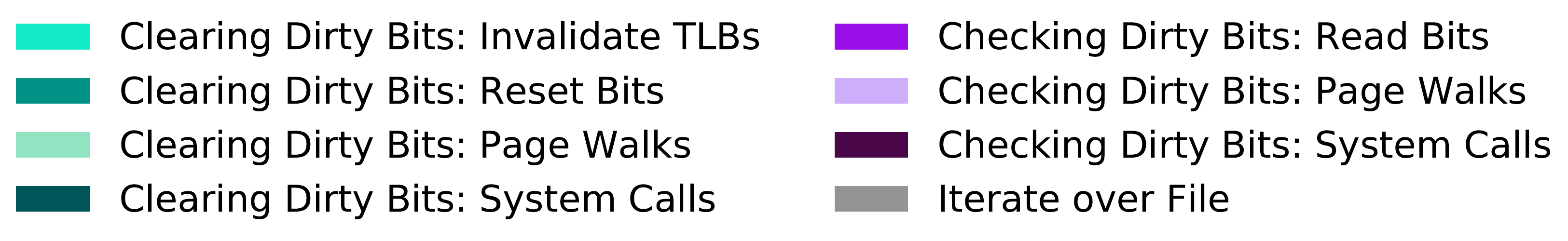}
	}
	\\[-1ex]
	 \begin{subfigmatrix}{2}
	    \subfigure[Breakdown of Time Spent]{\label{fig:breakdown}
			\includegraphics[width=0.45\columnwidth]{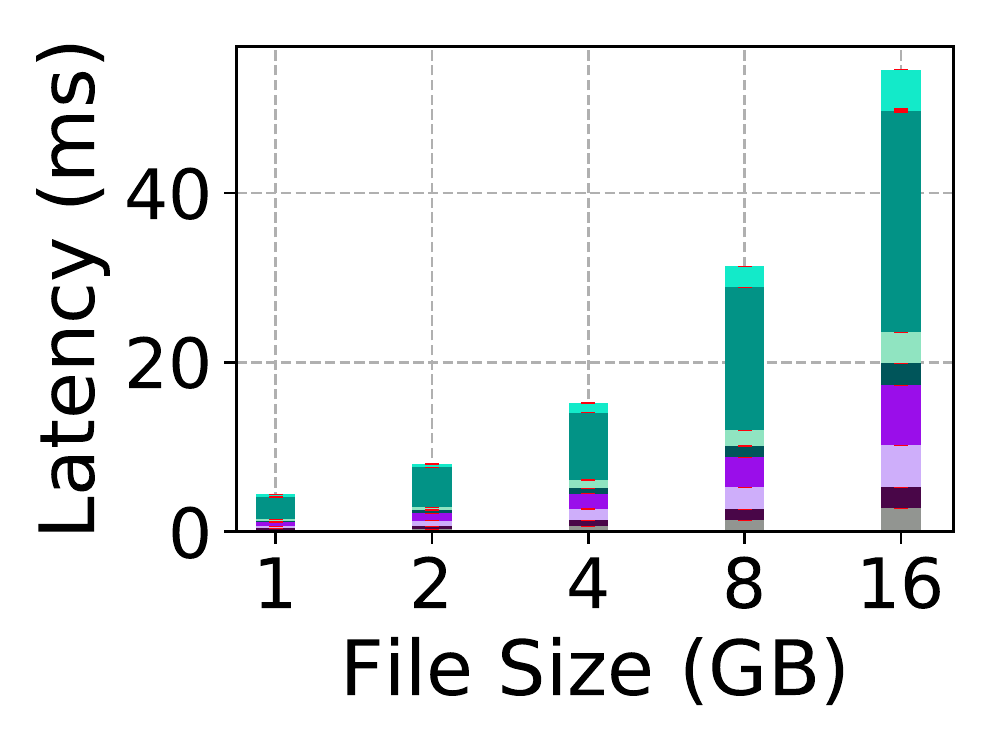}}
		\subfigure[Impact of Batch Size]{\label{fig:batch-size}
			\includegraphics[width=0.45\columnwidth]{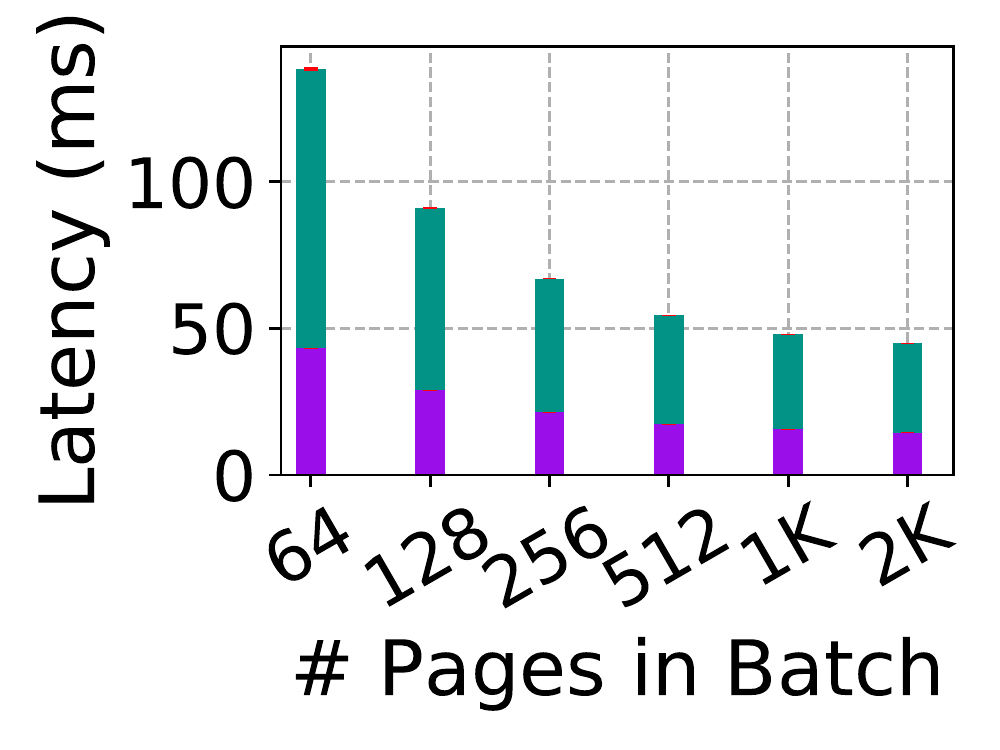}}
	 \end{subfigmatrix}
 \caption{\textbf{Cost of Checking/Clearing Dirty Bits} -- 
	 \ref{fig:breakdown} shows the time spent in each component 
	 of checking/clearing dirty bits  
	 for a batch size of 512 pages and increasing file sizes. 
	 \ref{fig:batch-size} shows that
	 increasing the batch size 
	 reduces the time spent in
	 checking/clearing dirty bits with diminishing returns.
	}
\label{fig:check-clear-dirty-bits}
\end{figure}

\textbf{Results}: \cref{fig:breakdown} 
presents the time spent in various components of checking and clearing 
dirty bits. The batch size
is set to 512 pages for this experiment. 
Doubling the file size, and consequently the total number of 
pages, roughly doubles the amount of time 
spent in each of the components. This is because 
the number of system calls, page walks, and reads of the 
dirty bits are all directly 
proportional to the total number of pages. The 
number of pages for which the dirty bit is cleared and 
the number of TLB invalidations
depend on the workload's access pattern. For the uniform 
random access workload, these are also 
directly proportional to the total number of pages.

\cref{fig:batch-size} presents the impact of batch size 
for a 16~GB file. As the batch size increases, 
the time spent in checking/clearing dirty bits decreases 
with diminishing marginal returns. 
This decrease is because the number of system 
calls reduce and larger fractions of the page table walks 
are shared between the pages in the same batch. 
The benefits are diminishing with increasing batch 
size, because of the 
fixed cost of reading all the dirty bits and resetting the ones 
that are found to be set.

\subsection{Battery Capacity Requirements}
\label{sec:battery}
This section analyzes the cost of batteries required for \system 
to update the \redundancy after a power failure for 
various workloads. 
We consider two kinds of batteries: ultra-capacitors 
that cost \$2.85/KJ~\cite{wsp-asplos, datacenter-energy-storage-sigmetrics}, 
and lithium-ion batteries 
that cost \$0.02/KJ~\cite{datacenter-energy-storage-sigmetrics, 
cdn-batteries-socc}. Conventionally, datacenters use 
lithium-ion batteries; modern datacenters additionally use 
ultra-capacitors because of their 
higher energy efficiency and density~\cite{datacenter-energy-storage-sigmetrics}.
We consider servers with  
500W~\cite{datacenter-energy-storage-sigmetrics} 
power usage. 

For Redis with the write-heavy 
workload YCSB-A, one iteration of \system's \redundancy 
updates takes 143~ms when performed every second and 
562~ms when performed every 10 seconds. These correspond 
to less than 1~KJ of energy required, 
i.e. the cost would be 
less than \$2.85 when using ultra-capacitors and 
less than \$0.02 when using the conventional 
lithium-ion batteries.
This is the case for all PMDK key-value stores except 
RTree as well. For RTree, because of its sparse and 
large writes, \system can require up to 5 seconds 
to update the \redundancy upon a power failure, 
requiring 
2.5~KJ of energy. 
This corresponds to 
\$7.2 in
ultra-capacitor cost or 
\$0.05 lithium-ion 
battery cost. 
For \fio, even with the adversarial random 
write workload with a \redundancy update period of 
every 60 seconds, \system requires only 4.5 seconds 
after a power failure. This translates to 
2.25~KJ of 
required energy 
and 
\$6.4 in ultra-capacitor cost or 
\$0.04 in lithium-ion battery cost. 
The battery requirement, and the associated 
cost, can be further reduced by limiting the number 
of pages that can be dirty (i.e., with outdated \redundancy) 
using Viyojit's~\cite{viyojit-isca} design.

%% file: reliability-eval.tex
\newcommand{\mttf}{MTTF\xspace}
\newcommand{\fraction}{f\xspace}
\newcommand{\mttdl}{MTTDL\xspace}
\newcommand{\mttdlbaseline}{MTTDL_{\baseline}\xspace}
\newcommand{\mttdlsystem}{MTTDL_{\system}\xspace}
\newcommand{\mdlr}{MDLR\xspace}
\newcommand{\mdlrbaseline}{MDLR_{\baseline}\xspace}
\newcommand{\mdlrsystem}{MDLR_{\system}\xspace}
\newcommand{\mttfpage}{MTTF_{\textsc{Page}}\xspace}
\newcommand{\numpages}{P\xspace}
\newcommand{\stripewidth}{N\xspace}
\newcommand{\dirtystripes}{V\xspace}

\subsection{Reliability Analysis}
\label{sec:reliability-analysis}
We now evaluate the increase in mean time to data loss 
(\mttdl) over \baseline when using \system.
For \baseline, 
a single page corruption causes data loss.
$\mttdlbaseline = \frac{\mttfpage}{\numpages}$, 
where \numpages is the number of pages in the system.

A page corruption affects data protected with \system 
in different ways. If the corruption affects a page that is dirty, 
\system would checksum the corruption, 
leading to a silent data corruption. If 
the corruption affects a page that is itself clean but 
belongs to a stripe with a dirty page (hence, an outdated 
parity), \system cannot recover the page, causing a data loss.
For a corruption that affects a page that 
is itself clean and belongs to a stripe with all clean pages, 
\system can recover the page. In summary, if the corruption affects 
a page in a \emph{vulnerable stripe}, i.e., a stripe with even 
one dirty page, it would lead to data loss. 
$\mttdlsystem = \frac{\mttfpage}{\dirtystripes \times \stripewidth}$, 
where \dirtystripes is the number of vulnerable stripes, and 
\stripewidth is the number of pages in a stripe. 
\system increases the $\mttdl$ by 
$\frac{\numpages}{\dirtystripes \times \stripewidth}$ 
in comparison to \baseline. 

We use the above to compute the increase in the \mttdl with \system
over \baseline for the various applications and workloads 
described in \cref{sec:eval}.
Workload access patterns, i.e., the rate and locality of 
their data updates determine the number of vulnerable 
stripes. We emperically measure the average number of vulnerable 
stripes for the various workloads and use that to 
compute the increase in \mttdl.
For Redis, \system with a \redundancy 
update period of 1~sec increases the 
\mttdl by 15$\times$ for the write-heavy 
workload YCSB-A and 74$\times$ for the ready-heavy 
workload YCSB-B. Increasing the delay 
reduces the \mttdl, because a larger fraction 
of data remains dirty (e.g., 21$\times$ and 13$\times$ 
for YCSB-B with 5~sec and 10~sec period, respectively). 
For PMDK's key-value stores, \system increases 
the \mttdl by up to two orders of magnitude 
(e.g., 112$\times$ for RBTree insert-only workload 
with 32 threads).

%% file: conclusion.tex
\section{Conclusion}
\label{sec:conclusion}
\system provides low-overhead \redundancy for 
DAX NVM data by embracing an asynchronous approach. 
In doing so, \system creates a tunable trade-off 
between performance and time-to-coverage. For example, 
decreasing the \redundancy update delay from 5 seconds 
to 1 second reduces \system's throughput for Redis with
YCSB-A workload by 10\% but 
also increases the MTTDL by 3$\times$.
\system's asynchronous approach amortizes the performance 
overhead of updating \redundancy over 
multiple data writes. As a result, \system 
outperforms the state-of-the-art synchronous 
\redundancy solution, Pangolin, by up to 5$\times$. 
Although \system's delayed data coverage design is not 
suited for all applications, it adds a high throughput 
option to the suite 
of DAX NVM \redundancy options available to applications.

